%% file: large_EJ_paper_arxiv.tex
\documentclass[aps,prb,twocolumn,floatfix,letterpaper,superscriptaddress,amssymb,eprint,longbibliography]{revtex4-2}
\usepackage{graphicx}
\usepackage[intlimits]{amsmath}

\allowdisplaybreaks[1]

\begin{document}

%\newcommand{\modified}[1]{\textcolor{black}{#1}}

%\title{A Superconducting Circuit as a Photon-Instanton Collider}
\title{Photon-instanton collider implemented by a superconducting circuit}

\author{Amir Burshtein}
\affiliation{Raymond and Beverly Sackler School of Physics and Astronomy, Tel Aviv University, Tel Aviv 6997801, Israel}

\author{Roman Kuzmin}
\affiliation{Department of Physics, University of Maryland, College Park, MD 20742, USA}
%\affiliation{Department of Physics, Joint Quantum Institute, and Center for Nanophysics and Advanced Materials, University of Maryland, College Park, MD 20742, USA}

\author{Vladimir E. Manucharyan}
\affiliation{Department of Physics, University of Maryland, College Park, MD 20742, USA}
%\affiliation{Department of Physics, Joint Quantum Institute, and Center for Nanophysics and Advanced Materials, University of Maryland, College Park, MD 20742, USA}

\author{Moshe Goldstein}
\affiliation{Raymond and Beverly Sackler School of Physics and Astronomy, Tel Aviv University, Tel Aviv 6997801, Israel}

\begin{abstract}
Instantons, spacetime-localized quantum field tunneling events, are ubiquitous in correlated condensed matter and high energy systems. However, their direct observation through collisions with conventional particles has not been considered possible. We show how recent advances in circuit quantum electrodynamics, specifically, the realization of galvanic coupling of a transmon qubit to a high-impedance transmission line, allows the observation of inelastic collisions of single microwave photons with instantons (phase slips). We develop a formalism for calculating the photon-instanton cross section, which should be useful in other quantum field theoretical contexts. %in both condensed matter and high energy systems.
In particular, we show that the inelastic scattering probability can significantly exceed the effect of conventional Josephson quartic anharmonicity, and reach order-unity values.
%, and show that such collisions result in strong inelastic scattering of single photons, exceeding other sources of nonlinearity.
\end{abstract}

\maketitle

%\newpage\pagebreak
%\begin{widetext}
%\appendix

%\section*{Supplemental Material}

\emph{Introduction.---}
Instantons are time-localized solutions to a system's imaginary time equations of motion, describing quantum tunneling events. They typically bridge between symmetry-related configurations and carry nontrivial topological indexes~\cite{marino}. Instantons play important roles in many areas of physics, ranging from single-particle quantum-mechanical tunneling~\cite{marino}, through transport in low dimensional superconductors and superfluids (where they are also known as ``phase slips'', and can be thought of as vortices crossing the system)~\cite{schon90,fazio01,leggett,hriscu11,rastelli13,ergul13,vogt16,bard17}, to determining the phase diagram~\cite{polyakov} and breaking of classical conservation laws~\cite{coleman,gross81} in gauge theories.
Most of these studies concern thermodynamic or transport properties. A more direct way to probe such short-lived excitations would be through resonances they may induce in the scattering cross sections or decay rates of other more stable particles with which they interact. However, %the question of the interaction and scattering of instantons with other excitations 
such questions received much less attention, in large part due to lack of relevant experiments.

Advances in the fabrication and control of superconducting circuits allow monitoring of the dynamics of single microwave photons. For example, recent experiments have exposed unusual relaxation dynamics in a uniform Josephson junction array, in which phase slips play an important role~\cite{kuzmin19a}. However, their interpretation is complicated due to the presence of disorder and offset charge fluctuations~\cite{bard18,houzet19,wu19}.
It has recently been realized theoretically~\cite{camalet04,garciaripoll08,lehur12,goldstein13,peropadre13,snyman15,gheeraert17,leppakangas18,gheeraert18,yamamoto19,belyansky20} that controllable quantum simulation of many-body physics may be easier to achieve in ``quantum impurity'' setups, leading to initial experiments~\cite{forndiaz17,magazzu18,kuzmin19b,puertasmartinez19,indrajeet20}.
We thus study a single flux-tunable small Josephson junction in the regime where the Josephson energy still dominates (transmon qubit~\cite{koch07}), galvanically coupled to an array of large junctions. The array acts as a transmission line allowing microwave photons to controllably scatter off the small junction~\cite{kuzmin19b}. The large Josephson inductance makes the line wave impedance of the order of the resistance quantum, hence the array screens the effects of unwanted offset charges on the transmon without completely suppressing phase slips there. From a broader perspective, the large impedance amounts to an effective fine-structure constant of order unity~\cite{manucharyan09}, ushering in unprecedentedly strong correlations.
%Superstrong transmon-array coupling has recently been demonstrated in this system~\cite{kuzmin19b}.
We will show that a single photon propagating along the array may excite a phase slip at the transmon and inelastically convert into lower-frequency photons with a high probability, significantly larger than the conversion probability due to the usual Josephson quartic nonlinearity~\cite{fn:DCB}; %\nocite{ingold92,peugeot20};
this effect could be measured via the resulting broadening of the array modes~\cite{kuzmin19a}.
%In this work we will show how this setup could be used to reach a regime of significant inelastic collisions of single microwave photons off phase slips in the transmon, which would dominate over other nonlinearities in the system, and be measured via the resulting broadening of the individual array modes~\cite{kuzmin19a}.
For this we develop an extension of the standard equilibrium instanton calculation~\cite{marino} to a scattering scenario, which could be useful in other fields. We will now outline its main ingredients, deferring some technical details to the supplemental material~\cite{SM}.

\begin{figure}
	\centering
	%	\includegraphics[width=0.49\columnwidth,height=!]{circuit_sketch}
	%%	\hspace{-0.05\columnwidth}
	%	\includegraphics[width=0.49\columnwidth,height=!]{simplified_circuit_sketch}
	\includegraphics[clip, trim = 0.15cm 0cm 0.15cm 0cm, width=0.49\columnwidth,height=!]{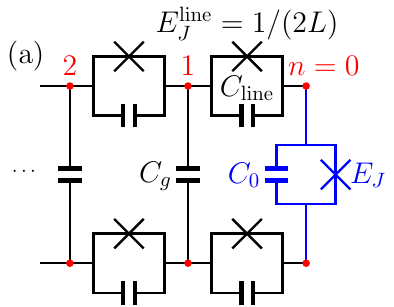}
%	\includegraphics[clip, trim = 0.1cm 0cm 0.1cm 0cm, width=0.49\columnwidth,height=!]{circuit_sketch.png}
	%\hspace{-0.03\columnwidth}
	\includegraphics[clip, trim = 0.15cm 0cm 0.15cm 0cm, width=0.49\columnwidth,height=!]{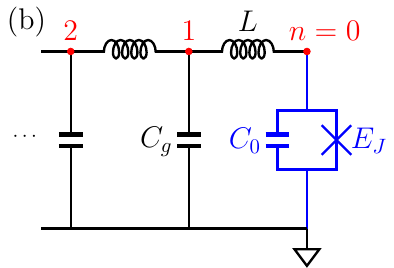}
	\caption{\label{fig:circuit}
		%%(Color online)
		The studied system: (a) The full circuit; (b) A simplified version. See the text for details.
	\vspace{-1.2cm}}
\end{figure}

\emph{Model.---}
We concentrate on the setup realized in a recent experiments~\cite{kuzmin19b,kuzmin20}, corresponding to the electric circuit depicted in Fig.~\ref{fig:circuit}(a). It consists of a long (length $N \gg 1$) two-leg array of superconducting islands connected by strong Josephson junctions $E_J^\mathrm{line}$ with large junction capacitance $C^\mathrm{line}$, negligible ground capacitance (not depicted), and intermediate inter-leg capacitance $C_g$. The large $C^\mathrm{line}$ suppresses phase slips along the arrays,  allowing their treatment as classical transmission lines. Except for this, $C^\mathrm{line}$ could be ignored below the array plasma frequency. The small ground capacitance pushes the leg-even modes to high frequencies, decoupling them from the transmon. %The inter-leg capacitance allows for linearly-dispersing leg-odd modes.
%These considerations allow us to
We may thus employ a simplified single-leg array model [Fig.~\ref{fig:circuit}(b)] for the leg-odd degrees of freedom.
The array capacitance to the ground $C_g$ and inductance $L$ in Fig.~\ref{fig:circuit}(b) are the inter-leg capacitance and twice the intra-leg Josephson inductance in Fig.~\ref{fig:circuit}(a), leading to a Lagrangian
%We start from the following Lagrangian, describing the odd modes of an $n$-junction double-leg superconducting ladder array terminated by a SQUID (neglecting junction capacitance, since we are interested in frequencies much smaller than the array plasma frequency):
\begin{equation} \label{eqn:lagrangian}
\mathcal{L} = \frac{C_0 \dot{\phi}_0^2}{2} + E_J \cos \left( 2\phi_0 \right) + \sum_{n=1}^{N} %\left[
\frac{C_g \dot{\phi}_n^2}{2} - \frac{\left(\phi_n - \phi_{n-1} \right)^2}{2L}, %\right],
\negthickspace
\end{equation}
where $\phi_n$ is in units of flux %(so that its derivative is the voltage),
and we employ units where $e=1$ and $\hbar = 1$, hence the flux quantum is $\Phi_0 = h/2e = \pi$. The array spacing $a$ will serve as the unit of length. %The array capacitance to the gate $C_g$ and inductance in Fig.~\ref{fig:circuit}(b) are the inter-leg capacitance and twice the intra-leg Josephson inductance in Fig.~\ref{fig:circuit}(a).

The array is terminated by a transmon qubit~\cite{koch07} (node $n=0$, blue elements in Fig.~\ref{fig:circuit}), a small SQUID whose Josephson energy $E_J$ is flux-tunable and much larger than its charging energy, $E_C = 1/2C_0$. Hence, to leading order we may approximate its Josephson cosine by a quadratic function~\cite{kuzmin19b}. %This is the situation realized in Ref.~\cite{kuzmin19b}.
Then Eq.~\eqref{eqn:lagrangian} gives rise to eigenmodes with dispersion $\omega_k = 2 v  \sin (k/2) \approx vk$, where $v=1/\sqrt{LC_g}$, the array wave velocity divided by the array spacing, is much larger than all other energy scales, i.e., for all relevant modes $k \ll \pi$.
The eigenmodes are $\propto \sin (k n +\delta_k)$, %with phase shifts $\delta_k = \tan^{-1}[\Gamma_0 \omega_k/(\omega_0^2 - \omega_k^2)]$.
where~\cite[][Sec.~SI.B]{SM} %the phase shifts $\delta_k$ obey
\begin{align} \label{eqn:dk}
%\sin \delta_k = & \frac{\Gamma_0 \omega_k}{\left[ \left( \omega_0^2 - \omega_k^2 \right)^2 + (\Gamma_0\omega_k)^2 \right]^{1/2}}, \nonumber \\
%\cos \delta_k = & \frac{ \omega_0^2 - \omega_k^2}{\left[ \left( \omega_0^2 - \omega_k^2 \right)^2 + (\Gamma_0\omega_k)^2 \right]^{1/2}}.
%\tan \delta_k = & \frac{\Gamma_0 \omega_k}{ \omega_0^2 - \omega_k^2}.
\delta_k = \tan^{-1} \left(\frac{\Gamma_0 \omega_k}{ \omega_0^2 - \omega_k^2}\right)
\end{align}
is the phase shift.
Here $\omega_0 = \sqrt{8 E_J E_C}$ is the trasmon LC frequency %(including leading order correction due to its anharmonicity),
and $\Gamma_0 = 1/Z C_0 = 4 E_C/\pi z$ is its elastic broadening due to its coupling to the array, where $Z=\sqrt{L/C_g}$ is the array wave impedance and $z=Z/R_Q$ ($R_Q = h/(2e)^2 = \pi/2$ is the superconducting resistance quantum).
%In the limit of a disconnected array, the allowed wavevectors obey $k N - \delta_k = \pi m$, $m \in \mathbb{Z}$, leading to a
For $N \gg 1$ the mode spacing is $\Delta = \pi v/N$, hence $\sum_k \to \int_0^\infty \mathrm{d}\omega/\Delta$.

Upon increasing $E_C/E_J$ the transmon nonlinearity starts becoming significant. We will concentrate on the regime where $\sqrt{E_C/E_J}$ is still small, and furthermore, $\Gamma_0/\omega_0 \ll 1$ (i.e., $\sqrt{E_C/E_J} \ll z$), so the transmon resonance is well-defined~\cite{kuzmin19b}. %Hence, $\sqrt{E_C/E_J} \ll \min(z,1)$.
In this regime the nonlinearity manifests in two ways: (i) Expanding the Josephson cosine gives rise to quartic nonlinearity, shifting $\omega_0$ by $-E_C$. It could also induce photon inelastic scattering, but we will show later on that for realistic device parameters this effect could be subleading; (ii) The periodicity of the cosine allows for instantons (phase slips). An incoming photon may excite a phase slip, and the resulting voltage and current pulse may give rise to the emission of photons with different frequencies.
We will now study in detail the latter inelastic effect. %collisions of single photons with them.

\emph{Instanton calculation.---}
For a disconnected transmon [first two terms of Eq.~\eqref{eqn:lagrangian}] %, or, equivalently, $z \to \infty$)
the classical instanton solution in imaginary time, describing a phase slip between $\phi_0 (\tau \to -\infty) = 0$ %at $\tau \to -\infty$
and $\phi_0 (\tau \to \infty) = \pm \Phi_0 = \pm \pi$, %$\tau \to \infty$,
is $\phi_0^{(0)} (\tau) = \pm 2 \tan^{-1} \left( e^{\omega_0\tau} \right)$, or, in Fourier space, $\phi_0^{(0)} (\omega) = \pm \pi / i\omega \cosh \left( \pi \omega/2 \omega_0 \right)$~\cite{marino}. Here and below, the upper (lower) sign corresponds to an instanton (anti-instanton).
%\begin{align} \label{eqn:phi0_tau}
%\phi_0^{(0)} (\tau) = 2 \tan^{-1} \left( e^{\omega_0\tau} \right).
%\end{align}
%Its Fourier transform is
%\begin{align}
%\phi_0^{(0)} (\omega) = \frac{\pi}{ i\omega \cosh \left( \frac{\pi}{2} \frac{\omega}{\omega_0} \right) }.
%\end{align} \label{eqn:phi0_omega}
The classical action $S_0$ of the instanton, together with the contributions of Gaussian fluctuations around it, give rise to the transmon ground state charge dispersion $\lambda_0$ (half the width of the lowest Bloch band of the corresponding Mathieu equation~\cite{koch07,abramowitz}) in the WKB approximation~\cite[][Sec. SI.A]{SM}, %~\cite{koch07} %, the half bandwidth of the lowest Bloch band of the corresponding Mathieu equation~\cite{koch07,abramowitz}. In the WKB approximation,
\begin{align} \label{eqn:l0}
\lambda_0 \approx \frac{8}{\sqrt{\pi}} \left( 8 E_J^3 E_C \right)^{1/4} e^{-\sqrt{8E_J/E_C}}.
\end{align}
%More accurate expression for $\lambda_0$ can be written in terms of the Mathieu characteristic values~\cite{koch07,abramowitz,SM}.
%Instanton effects will become nonperturbative for $z<1$ and frequencies below the renormalized $\lambda_0$, hence below we will assume $\Gamma_0$ or $T$ to be larger than this scale.

We now incorporate the array %effect of the array on a phase slip
to lowest order in $\Gamma_0/\omega_0$. Expanding the imaginary time action around the classical isolated instanton solution [$\phi_0^{(0)}(\tau)$ as given above and $\phi_{n > 0}^{(0)}(\tau)=0$] to second order in the deviation $\delta\phi_n = \sum_k \delta\phi_k \sin(k n+\delta_k)$, one finds~\cite[][Sec.~SI.B]{SM}:
\begin{align} \label{eqn:ds}
S = S_0 & + \int \frac{\mathrm{d} \omega}{2\pi}
\left[
\frac{\left|\phi_0^{(0)} (\omega)\right|^2}{2L}
+  \sum_k %\left[
\frac{C_k}{2} (\omega^2+\omega_k^2) \left| \delta\phi_k(\omega) \right|^2
\right.%\right.
\nonumber \\
& \qquad \quad
\left.%\left.
- \frac{\sin(k+\delta_k) - \sin(\delta_k)}{L} \phi_0^{(0)} (-\omega) \delta \phi_k (\omega) \right]
%\right\}
\nonumber \\
& %\negthickspace %\negthickspace %\negthickspace %\negthickspace
- \int \mathrm{d}\tau %\left\{%\sum_{k,k^\prime}
\frac{8 E_J}{\cosh^2 (\omega_0\tau)} \left[ \sum_k \sin(\delta_k)  \delta \phi_k(\tau) \right]^2, %\right\}, %\sin(\delta_{k^\prime}) \delta \phi_{k^\prime}(\tau).
\end{align}
where the capacitance of mode $k$ is $C_k \approx N C_g/2$ for $N \gg 1$. The very last term contributes to higher orders in $\Gamma_0/\omega_0$ and will be neglected henceforth. The classical equations of motion for $\delta \phi_k$ result in
\begin{align} \label{eqn:dphik}
\delta \phi_k(\omega)
%& = \frac{1}{C_k (\omega^2 + \omega_k^2)} \frac{\sin(k-\delta_k) + \sin(\delta_k)}{L} \phi_0^{(0)} (\omega)
%\nonumber \\ &
\approx \frac{1}{C_k (\omega^2 + \omega_k^2)} \frac{\omega_k \cos\delta_k}{Z} \phi_0^{(0)} (\omega) ,
\end{align}
to leading order in $k \ll 1$. %where we assume that %the last approximation is valid when
%for the relevant modes $k \ll 1$, that is, $\omega_0,\omega_k \ll v$.
Plugging this back into the action~\cite{fn:fluctuations} gives~\cite[][Sec.~SI.B]{SM} %a contribution to the classical instanton action
%~\footnote{Corrections to the fluctuations contribution are subleading in $\Gamma_0/\omega_0$}:
\begin{align} \label{eqn:fk_tilde}
\delta S = \frac{1}{2} \sum_k \tilde{f}_k^2,
%\end{align}
%with $\tilde{f}_k$ given by
%\begin{align} \label{eqn:fk_tilde}
\quad
\tilde{f}_k = \sqrt{\frac{2\Delta}{z\omega_k}}
\frac{1}{\mathrm{cosh} \left(\frac{\pi}{2}\frac{\omega_k}{\omega_0} \right)},
\end{align}
leading to a renormalization $\lambda_0 \to \lambda_0 e^{-\sum_k \tilde{f}_k^2/2}$. For $z>1$ instantons are relevant, resulting in an emergent scale, $\lambda_* \sim \lambda_0 (\lambda_0/\omega_0)^{1/(z-1)}$, below which instanton effects are nonperturbative~\cite{schon90}; we limit ourselves to higher energies.
% , in particular $\lambda_* \ll \max(\Gamma_0,T)$.

Within the approximations we employ, the contribution of a single instanton to a multipoint correlation of the $\phi_k$ is given by the corresponding classical solution~\cite{fn:fluctuations}, multiplied by %the instanton action+fluctuations contribution,
$\lambda_0 e^{-\sum_k \tilde{f}_k^2/2}/2$. By the %Lehmann–Symanzik–Zimmermann
LSZ reduction formula~\cite{peskin,zhou96}, this correlation with its external single-particle legs amputated gives the $\mathcal{T}$-matrix element between $N_\mathrm{in}$ incoming photons with momenta $k^\prime_1, k^\prime_2, \cdots, k^\prime_{N_\mathrm{in}}$ and $N_\mathrm{out}$ outgoing photons with momenta $k_1, k_2, \cdots, k_{N_\mathrm{out}}$~\cite[][Sec.~SI.C]{SM}:
\begin{widetext}
	\vspace{-12pt}
\begin{align} \label{eqn:T}
\mathcal{T}^{ k^\prime_1, k^\prime_2, \cdots, k^\prime_{N_\mathrm{in}} }_{ k_1, k_2, \cdots, k_{N_\mathrm{out}} } & =
\frac{\Delta}{2\pi}
%\lim_{\substack{ \omega^\prime_j \to i \omega_{k^\prime_j} \\ \omega_j \to -i \omega_{k_j} }
\lim_{\substack{ \{ \omega^\prime_j \to i \omega_{k^\prime_j} \} \\ \{ \omega_j \to -i \omega_{k_j} \} }
}
\prod_{j=1}^{N_\mathrm{in}}
\frac{C_{k^\prime_j} \left( \omega^{\prime 2}_j  + \omega_{k^\prime_j}^2 \right) }{ \sqrt{2 C_{k^\prime_j} \omega_{k^\prime_j} } }
\prod_{j=1}^{N_\mathrm{out}}
\frac{C_{k_j} \left( \omega^{ 2}_j  + \omega_{k_j}^2 \right) }{ \sqrt{2 C_{k_j} \omega_{k_j} } }
%\times \nonumber \\ & \qquad \qquad \qquad
\left\langle
\prod_{j=1}^{N_\mathrm{in}}
\phi_{k^\prime_j}(\omega^\prime_j)
\prod_{j=1}^{N_\mathrm{out}}
\phi_{k_j}(\omega_j)
\right\rangle _\mathrm{1-instanton}
%\Big|
%^{ \omega^\prime_i \to i \omega_{k^\prime_i}}
%_{ \substack{ \omega_i \to -i \omega_{k_i} \\ \mathrm{1-instanton}}}
\nonumber \\ &
= (\mp 1)^{N_\text{in}} (\pm 1)^{N_\text{out}} f_{k^\prime_1} f_{k^\prime_2} \cdots f_{ k^\prime_{N_\mathrm{in}} }
f_{k_1} f_{k_2} \cdots f_{ k_{N_\mathrm{out}} }
\frac{\lambda_0}{2} e^{-\sum_k \tilde{f}_k^2/2}
%\mathcal{T}^{ k^\prime_1, k^\prime_2, \cdots, k^\prime_{N_\mathrm{in}} }_{ k_1, k_2, \cdots, k_{N_\mathrm{out}} } & =
%\frac{\Delta}{2\pi}
%\frac{C_{k^\prime_1} \left( \omega^{\prime 2}_1  + \omega_{k^\prime_1}^2 \right) }{ \sqrt{2 C_{k^\prime_1} \omega_{k^\prime_1} } }
%\cdots
%\frac{C_{k^\prime_{N_\mathrm{in}}} \left( \omega^{\prime 2}_{N_\mathrm{in}}  + \omega_{k^\prime_{N_\mathrm{in}}}^2 \right) }{ \sqrt{2 C_{k^\prime_{N_\mathrm{in}}} \omega_{k^\prime_{N_\mathrm{in}}} } }
%%
%\frac{C_{k_1} \left( \omega^{ 2}_1  + \omega_{k_1}^2 \right) }{ \sqrt{2 C_{k_1} \omega_{k_1} } }
%\cdots
%\frac{C_{k_{N_\mathrm{out}}} \left( \omega^{ 2}_{N_\mathrm{out}}  + \omega_{k_{N_\mathrm{out}}}^2 \right) }{ \sqrt{2 C_{k_{N_\mathrm{out}}} \omega_{k_{N_\mathrm{out}}} } }
%\times \nonumber \\ & \qquad \qquad \qquad
%\left\langle
%\phi_{k^\prime_1}(\omega^\prime_1) \cdots \phi_{k^\prime_{N_\mathrm{in}}} (\omega^\prime_{N_\mathrm{in}})
%\phi_{k_1}(\omega_1) \cdots \phi_{k_{N_\mathrm{out}}} (\omega_{N_\mathrm{out}})
%\right\rangle_\mathrm{1-instanton}
%\Big|
%^{ \omega^\prime_1 \to i \omega_{k^\prime_1}, \cdots, \omega^\prime_{N_\mathrm{in}} \to i \omega_{k^\prime_{N_\mathrm{in}}} }
%_{ \omega_1 \to -i \omega_{k_1}, \cdots, \omega_{N_\mathrm{out}} \to -i \omega_{k_{N_\mathrm{out}}} }
%\nonumber \\ &
%= (\mp 1)^{N_\text{in}} (\pm 1)^{N_\text{out}} f_{k^\prime_1} f_{k^\prime_2} \cdots f_{ k^\prime_{N_\mathrm{in}} }
%f_{k_1} f_{k_2} \cdots f_{ k_{N_\mathrm{out}} }
%\frac{\lambda_0}{2} e^{-\sum_k \tilde{f}_k^2/2}
\end{align}
\end{widetext}
with %the upper (lower) sign corresponding to an instanton (anti-instanton), and
\begin{align} \label{eqn:fk_inst}
%\negthickspace\negthickspace\negthickspace %\negthickspace
f_k = \sqrt{\frac{2\Delta}{z\omega_k}}
\frac{ \omega_0^2 - \omega_k^2 }{ \sin \left(\frac{\pi}{2}\frac{\omega_0-\omega_k}{\omega_0} \right) \sqrt{\left( \omega_0^2 - \omega_k^2 \right)^2 + (\Gamma_0 \omega_k)^2}},
%\frac{ \omega_0^2 - \omega_k^2 }{ \left[ \left( \omega_0^2 - \omega_k^2 \right)^2 + (\Gamma_0 \omega_k)^2 \right]^{1/2} \negthickspace \sin \left(\frac{\pi}{2}\frac{\omega_0-\omega_k}{\omega_0} \right)},
%f_k = \sqrt{\frac{2\Delta}{z\omega_k}}
%\frac{ \left(\omega_0^2 - \omega_k^2\right) \sec \left(\frac{\pi}{2}\frac{\omega_k}{\omega_0} \right) }{ \left[ \left( %\omega_0^2 - \omega_k^2 \right)^2 + (\Gamma_0 \omega_k)^2 \right]^{1/2} },
\end{align}
being the ``form factor'' of the instanton in the photon modes basis.
Note that it is finite at the resonance frequency $\omega_0$ but still peaked there. It rises toward low frequencies (assumed higher than $\lambda_*$); this reflects the fact that an instanton involves a shift of phases along the entire array, and hence couples well to low-$k$ modes.
Thus, processes in which a nearly resonant photon scatters into one nearly resonant photon and several low energy photons (whose number is controlled by $z$) will play an important role.
Note also that $f_k$ diverges at higher odd multiples of $\omega_0$, which are nonlinear resonances broadened only at higher order in $\Gamma_0$. In relevant experiments~\cite{kuzmin19b} these will anyway be close to $E_J$, i.e., outside the instanton regime, hence we will limit ourselves here to lower frequencies. Adding up the contribution of the instantons and the anti-instantons eliminates processes involving an odd number of photons.

\begin{figure*}
	\centering
	\includegraphics[width=0.7\columnwidth,height=!]{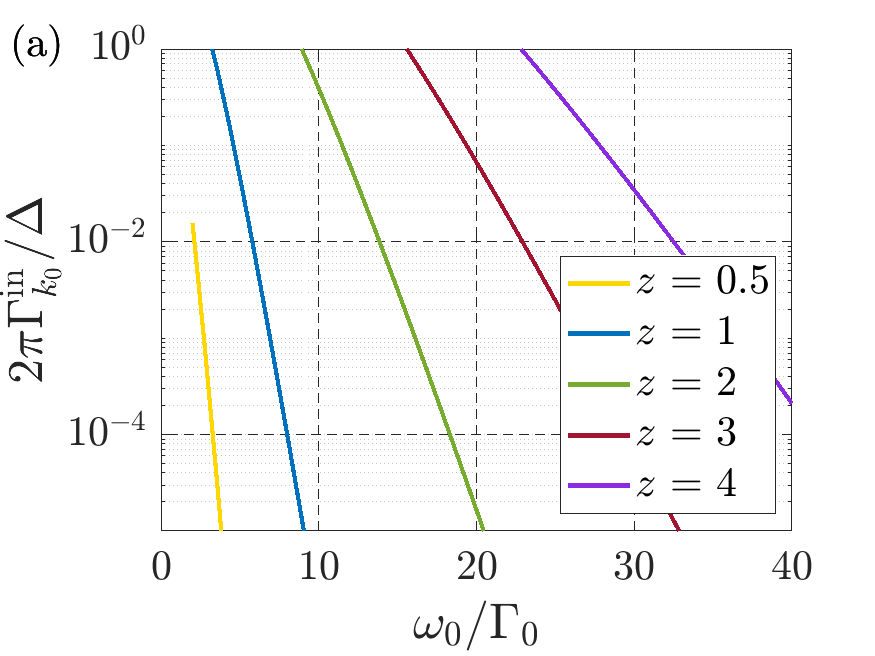}
	\hspace{-0.05\columnwidth}
	\includegraphics[width=0.7\columnwidth,height=!]{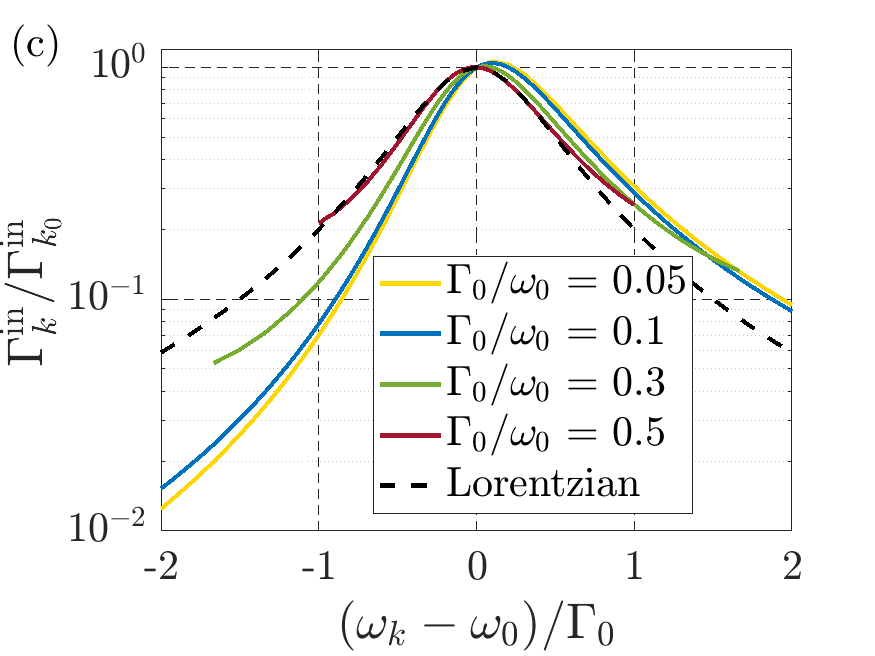}
	\hspace{-0.05\columnwidth}
	\includegraphics[width=0.7\columnwidth,height=!]{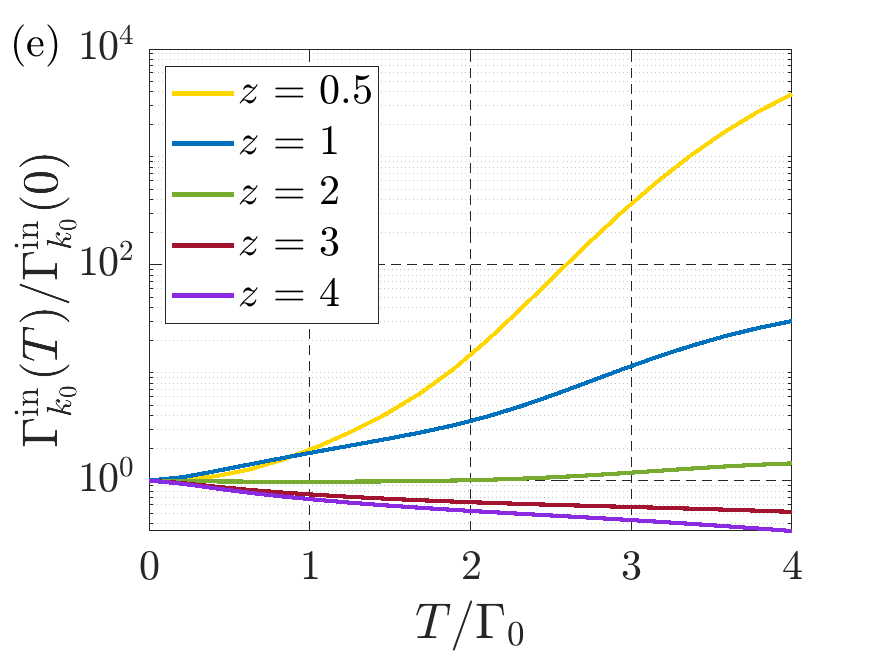}
	\newline
	\includegraphics[width=0.7\columnwidth,height=!]{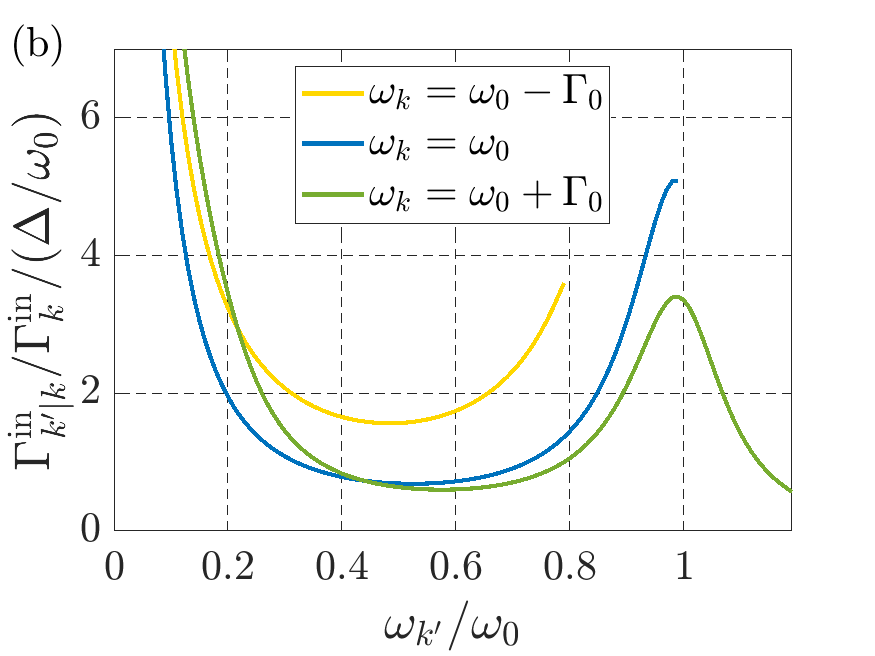} %{Gamma_inel_on_resonance_mathieu_wout_lambda_0.png}
	\hspace{-0.05\columnwidth}
	\includegraphics[width=0.7\columnwidth,height=!]{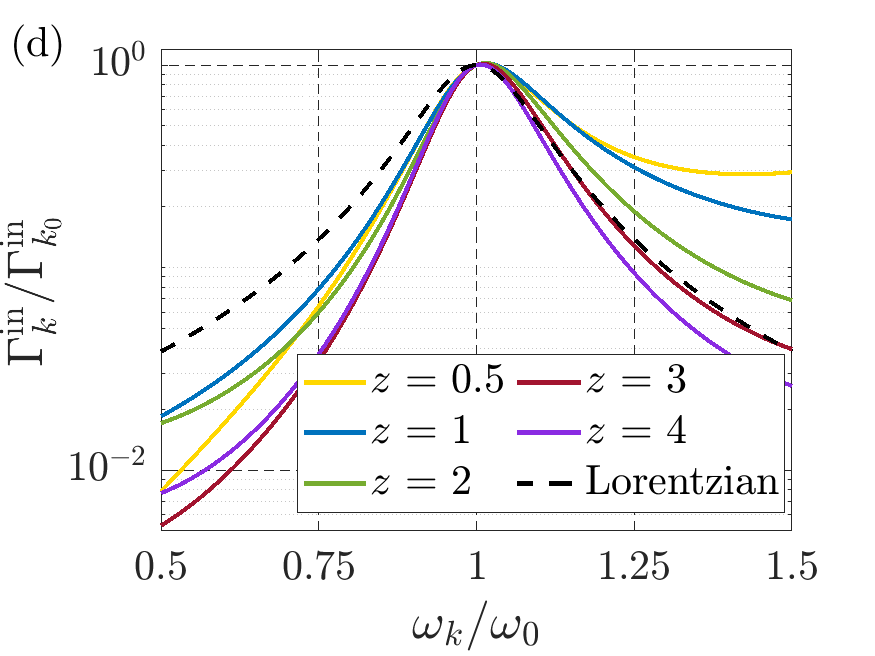}
	\hspace{-0.05\columnwidth}
	\includegraphics[width=0.7\columnwidth,height=!]{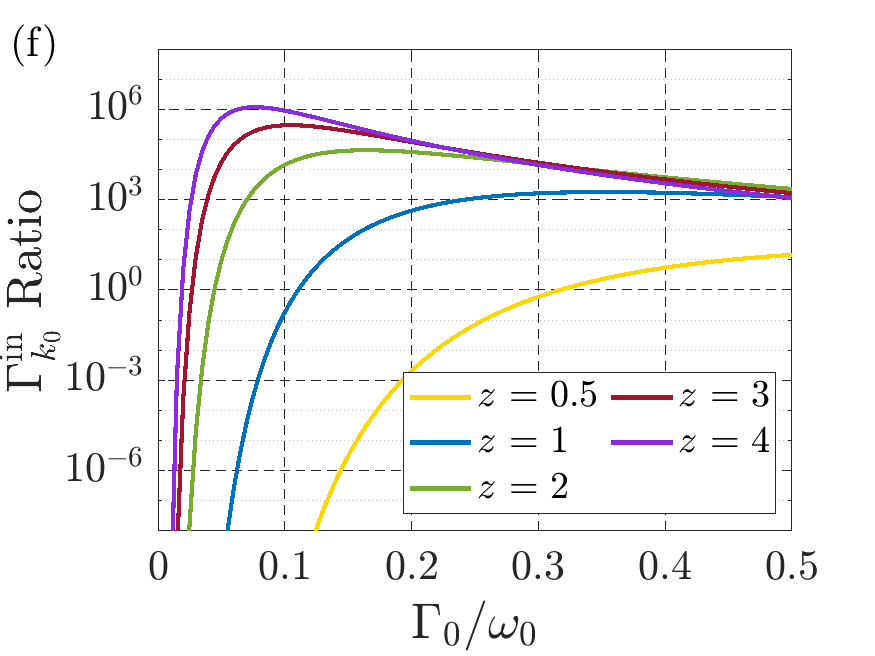}
	\vspace{-0.4cm}
	\caption{\label{fig:inelastic_rate}
		%%(Color online)
		Parameter dependence of the inelastic scattering probability %of $2\pi\Gamma^\mathrm{in}_k/\Delta$, the total inelastic scattering probability
		of a single incoming photon with frequency $\omega_k$ by a single phase slip, Eqs.~\eqref{eqn:pi_inst}--\eqref{eqn:ginel_kkp}.
		(a) On-resonance (mode $k_0 = \omega_0/v$) total probability $2\pi\Gamma^\mathrm{in}_{k_0}/\Delta$ as function of $\omega_0/\Gamma_0$ for several values of $z$ at $T=0$
		%(a) presents the total rate
		\{using the full Mathieu expression for $\lambda_0$~\cite{koch07,abramowitz}, rather than the approximate Eq.~\eqref{eqn:l0}~\cite[][Sec.~SI.A]{SM}\}.
		%(b) excludes the prefactor $\lambda_0^2$.
		(b) The distribution of inelastically generated photons $\Gamma^\mathrm{in}_{k^\prime|k}/\Gamma^\mathrm{in}_{k}$ at $\omega_k=\omega_0, \omega_0 \pm \Gamma_0$ for $z=2$ and $\Gamma_0/\omega_0=0.2$.
		(c,d) $T=0$ resonance lineshape at (c) $z=2$ and different $\Gamma_0/\omega_0$ or (d) $\Gamma_0/\omega_0=0.2$ and different $z$. A simple Lorentzian with width $\Gamma_0$ is also plotted for comparison.
		(e) Temperature dependence of the on-resonance probability for $\Gamma_0/\omega_0=0.05$ and different $z$. %($T \ll \omega_0$).
		(f)~Ratio between the $T=0$ on-resonance probabilities due to the instantons and due to the quatric nonlinearity, Eq.~\eqref{eqn:ginel_quartic}, showing that the former may dominate by several orders of magnitude for not-too-small $\Gamma_0/\omega_0$ and $z \gtrsim 1$.
		%by the instanton calculation and either the dual cosine approximation (continuous lines) or the limiting expression, Eq.~\eqref{eqn:2f1} (dashed lines).
		%See the text for further details. 
		%(c) uses fixed $z=2$, (e) uses fixed $\Gamma_0 = 0.05\omega_0$, and all other figures use a fixed value of $\Gamma_0 = 0.2\omega_0$
		\vspace{-0.6cm}
	}
\end{figure*}

Let us consider processes in which an additional photon with the specific frequency $\omega_k$ is included among either the incoming or outgoing photons. Combining the square of the $\mathcal{T}$-matrix elements just obtained with the appropriate Bose-Einstein factors corresponding to spontaneous and stimulated emission as well as stimulated absorption, gives the total rate $\Gamma^\mathrm{in}_k$ of the inelastic decay (minus creation) of a single incoming photon at $k$~\cite[][Sec.~SI.D]{SM}
(the inelastic scattering probability per collision is $2\pi\Gamma^\mathrm{in}_k/\Delta$, while $\omega_k/\Gamma^\mathrm{in}_k$ is the experimentally-measurable quality factor of mode $k$~\cite{kuzmin19b})~\cite{fn:thermodynamic},
%, which is finite in the thermodynamic limit),
\begin{widetext}
\begin{align} \label{eqn:ginel2_inst}
\Gamma^\mathrm{in}_k = &
\frac{\lambda_0^2}{2} f_k^2 e^{-\sum_{k^\prime} \tilde{f}_{k^\prime}^2  - 2 \sum_{k^\prime} f_{k^\prime}^2 n_B(\omega_{k^\prime})}
\sum_{N_\mathrm{out},N_\mathrm{in}}
\sum_{\substack{k_1<\cdots<k_{N_\mathrm{out}},\\ k_1^\prime<\cdots<k^\prime_{N_\mathrm{in}}}}
f_{k_1}^2 \cdots f_{k_{N_\mathrm{out}}}^2 f_{k_1^\prime}^2 \cdots f_{k^\prime_{N_\mathrm{in}}}^2
(1+n_B (\omega_{k_1})) \cdots (1+n_B (\omega_{k_{N_\mathrm{out}}}))
\times \nonumber
\\ %\nonumber
&	
%\negthickspace\negthickspace\negthickspace\negthickspace
%(1+n_B (\omega_{k_1})) \cdots (1+n_B (\omega_{k_{N_\mathrm{out}}}))
\qquad\qquad
n_B (\omega_{k^\prime_{1}}) \cdots n_B (\omega_{k^\prime_{N_\mathrm{in}}}) 
2\pi \left[ \delta \left( \omega_k + \omega_{k^\prime_1} + \cdots + \omega_{k^\prime_{N_\mathrm{in}}} -\omega_{k_1} - \cdots -\omega_{k_{N_\mathrm{out}}} \right)  - \{\omega_k \to -\omega_k\} \right],
\end{align}
%\end{widetext}
%where the sum is restricted to odd $N_\mathrm{out} + N_\mathrm{in} \ge 3$; however, in the thermodynamic limit one may discard this restriction and introduce a factor of half~\cite{SM}.
%The %term including $n_B(\omega_{k^\prime})$ in the exponent,
%factor $e^{- 2 \sum_{k^\prime} f_{k^\prime}^2 n_B(\omega_{k^\prime})}$, when expanded in a Taylor series~\cite{fn:thermodynamic}, %(assuming the thermodynamic limit $N \gg 1$, so that each $f_k^2 \sim 1/N$ is small, and one may ignore terms in which a single $k$ appears more than once),
%can be seen to account for the reduction in the probability of a process not involving photons with frequency $\omega_{k^\prime}$ when such photons are present, due to the increased probability of their emission or absorption.
%The term including $n_B(\omega_{k^\prime})$ in the exponent,
The probability of a process not involving photons with frequency $\omega_{k^\prime}$ decreases when such photons are present, due to the increased probability of their emission or absorption.
This is accounted for by the factor $e^{- 2 \sum_{k^\prime} f_{k^\prime}^2 n_B(\omega_{k^\prime})}$~\cite[][Sec.~SI.D]{SM}. %, as can be seen by expanding it in a Taylor series~\cite{SM}. %~\cite{fn:thermodynamic}.

%The probabilities for absorption and emission differ by a factor $e^{-\omega_k/T}$, hence difference between the two delta functions in the last expressions can be converted into their sum, up to a prefactor of $\tanh(\omega_k/2T)$, and
Upon expressing the delta functions via their Fourier representations, the summations over $N_\mathrm{in,out}$ and the $k$s can be recognized as the Taylor series of an exponent. %hyperbolic sine (without the linear term)~\cite{fn:thermodynamic},
%\footnote{In the thermodynamic limit $N \gg 1$, each $f_k^2 \sim 1/N$ is small, and one may ignore terms in which a single $k$ appears more than once.},
%which, for $\Delta \to 0$, can be approximated by an exponential function.
All in all we find that %the following rather compact expression for the inelastic rate
$\Gamma^\mathrm{in}_k = 2f_k^2 \Im \Pi_R (\omega_k)$, where
%\begin{widetext}
\begin{equation} \label{eqn:pi_inst}
\Pi_R(\omega) = -\lambda_0^2 %\tanh\frac{\omega}{2T}
%\left[
\int_0^\infty \mathrm{d}t
\sin(\omega t)
\exp \left(
-\sum_{k^\prime} \left\{  f_{k^\prime}^2 \left[ (1+n_B (\omega_{k^\prime}) ) (1 - e^{-i\omega_{k^\prime}t}) + n_B (\omega_{k^\prime}) (1 - e^{i\omega_{k^\prime}t})\right] + \tilde{f}_{k^\prime}^2 - f_{k^\prime}^2 \right\} \right), %\right],
\end{equation}
%\begin{equation} \label{eqn:ginel_inst}
%\Gamma^\mathrm{in}_k = 2\lambda_0^2 f_k^2 \tanh\frac{\omega_k}{2T} \mathrm{Re} \left[ \int_0^\infty \mathrm{d}t
%\cos(\omega_k t)
%\exp \left(
%-\sum_{k^\prime} \left\{  f_{k^\prime}^2 \left[ (1+n_B (\omega_{k^\prime}) ) (1 - e^{i\omega_{k^\prime}t}) + n_B (\omega_{k^\prime}) (1 - e^{-i\omega_{k^\prime}t})\right] + \tilde{f}_{k^\prime}^2 - f_{k^\prime}^2 \right\} \right)  \right].
%\end{equation}
%\end{widetext}
is the photon retarded self energy, whose imaginary part gives the total inelastic conversion (absorption minus emission) rate of energy $\omega$ into any photon combination. Using it, one may write down more refined rates; for example, the net rate of creation of photons at $k^\prime$ due to processes involving an incoming photon at $k$ is
%\begin{widetext}
%\begin{align}
%  \Gamma^\mathrm{in}_{k^\prime | k} = f_k^2 f_{k^\prime}^2 \left[
%  \Pi^{\prime\prime}_R \left( \omega_k - \omega_{k^\prime} \right) \left( \coth \frac{\omega_k - \omega_{k^\prime}}{2T}  + \coth \frac{\omega_{k^\prime}}{2T} \right)
%  + \Pi^{\prime\prime}_R \left( \omega_k + \omega_{k^\prime} \right) \left( \coth \frac{\omega_k + \omega_{k^\prime}}{2T}  - \coth \frac{\omega_{k^\prime}}{2T} \right)
%  \right],
%\end{align}
\begin{align} \label{eqn:ginel_kkp}
  \Gamma^\mathrm{in}_{k^\prime | k} = 2f_k^2 f_{k^\prime}^2 & \left\{
  \Im \Pi_R \left( \omega_k - \omega_{k^\prime} \right)
  \left[ (1+n_B (\omega_{k^\prime})) (1+n_B (\omega_k - \omega_{k^\prime}))
  - n_B (\omega_{k^\prime}) n_B (\omega_k - \omega_{k^\prime}) \right]
  %\left( \coth \frac{\omega_k - \omega_{k^\prime}}{2T}  + \coth \frac{\omega_{k^\prime}}{2T} \right)
  \right. \nonumber \\ & \left.
  + \Im \Pi_R \left( \omega_k + \omega_{k^\prime} \right)
  \left[ (1+n_B (\omega_{k^\prime})) n_B (\omega_k + \omega_{k^\prime})
  - n_B (\omega_{k^\prime}) (1+n_B (\omega_k + \omega_{k^\prime})) \right]
  %\left( \coth \frac{\omega_k + \omega_{k^\prime}}{2T}  - \coth \frac{\omega_{k^\prime}}{2T} \right)
  \right\},
\end{align}
\end{widetext}
which accounts for processes in which photons at $k,k^\prime$ are, respectively, absorbed-emitted, emitted-absorbed, emitted-emitted, or absorbed-absorbed, with appropriate signs to obey an energy conservation sum rule, $\omega_k \Gamma^\mathrm{in}_k = \sum_{k^\prime} \omega_{k^\prime} \Gamma^\mathrm{in}_{k^\prime | k}$.
The last couple of equations are the central results of this work. To recap, they apply for any $\omega_k, \omega_{k^\prime}$ between $\lambda_*$ and $3\omega_0$, provided that $\lambda_* \ll \max(\Gamma_0,T) \ll \omega_0$ and $E_C \ll E_J$. The single-instanton approximation further requires $2\pi\Gamma^\mathrm{in}_k/\Delta \lesssim 1$.

\emph{Inelastic rate behavior.---}
We exemplify the parameter dependence of the inelastic rate in Fig.~\ref{fig:inelastic_rate}.
To better understand its behavior, it is useful to study some limits~\cite[][Sec.~SI.E]{SM}.
%It is dominated by an inelastic resonance at $\omega_0$. Some of its salient features are: (i) The inelastic scattering probability can approach order unity. The charge dispersion $\lambda_0$ decreases fast with $\omega_0$, masking the corresponding increase in the number of possible decay channels contributing on-resonance [Fig.~\ref{fig:inelastic_rate}(a,b)]; (ii) The latter increase is visible in an asymmetry of the inelastic resonance lineshape, especially for small $\Gamma_0/\omega_0$ and $z$ [Fig.~\ref{fig:inelastic_rate}(c,d)]; (iii) Temperature suppresses coherent quantum phase slips (particularly for $z>1$, when they are relevant~\cite{schon90}) but gives rise to scattering by thermal photons, hence could either decrease or increase the decay rate, depending on $z$ [Fig.~\ref{fig:inelastic_rate}(e)].
%\emph{Limiting cases.---}
%The general expressions given above can be simplified further for nearly-resonant incoming photons if in addition $\Gamma_0/\omega_0 \to 0$ and $z > 1$~\cite{SM}. The energy sum rule is then dominated by photons with $\omega_{k^\prime}$ close to $\omega_0$ as well.
%At $T = 0$ the remaining photons share an energy $\sim \Gamma_0 \ll \omega_0$, for which $f_q~\sim \sqrt{2\Delta/z\omega_q}$ and thus~\cite{gogolin}
First, at $T=0$ and low frequencies $\omega \ll \omega_0$ one may approximate $f_q\approx\sim \sqrt{2\Delta/z\omega_q}$ in Eq.~\eqref{eqn:pi_inst}, leading to~\cite{gogolin}
\begin{align} \label{eqn:pi_inst_low}
  \Pi_R(\omega) \approx \frac{\pi}{\Gamma(2/z)} \frac{\lambda_0^2}{\omega}\left( \frac{\omega}{\omega_c(z)} \right)^{2/z},
  %\quad \omega \ll \omega_0,
\end{align}
where $\Gamma(x)$ is the gamma function~\cite{abramowitz}, and the effective cutoff $\omega_c(z) \approx 0.9 \omega_0$ is $z$ independent for $z \gtrsim 1$.

Let us now turn to the scattering of nearly-resonant photons, $\omega_k \approx \omega_0$, starting with $T=0$. As the spectrum of inelastically-emitted photons in Fig.~\ref{fig:inelastic_rate}(b) exemplifies, for $z\gtrsim 1$ and $\Gamma_0/\omega_0 \to 0$ the dominant process involves one emitted photon at $\omega_{k^\prime} \approx \omega_k$, while the other photons carry low energy of order $\Gamma_0$, hence
\begin{equation} \label{eqn:ginel_kkp_knk0}
	\frac{ 2\pi\Gamma^\mathrm{in}_{k^\prime | k} }{\Delta^2}
	\negthickspace \approx \negthinspace
	\frac{2 \lambda_1^2}{\Gamma(2/z) \omega_c}
	\negthinspace
	\left(\negthinspace
	\frac{\omega_k - \omega_{k^\prime}}{\omega_c}
	\negthinspace\right)^{\negthickspace\negthinspace\negthickspace\frac{2-z}{z}}
%	\times \nonumber \\
	\negthickspace\negthinspace\negthickspace\negthinspace\negthickspace
	\prod_{q=k,k^\prime}
	\negthickspace
	\frac{\frac{\Gamma_0}{2}}{(\omega_0 - \omega_{q})^2 + \left(\frac{\Gamma_0}{2}\right)^2},
\end{equation}
where $\lambda_1 = -\sqrt{2^7 E_J/E_C} \lambda_0$ is the charge dispersion of the first excited level of an isolated transmon~\cite{koch07}.
%For $\Gamma_0 \lesssim \omega_0-\omega_c(z)$,
Summing over $k^\prime$ one obtains on resonance (mode $k_0 = \omega_0/v$),
%\begin{align} \label{eqn:2f1}
%\negthickspace\negthickspace
%  \frac{ 2\pi\Gamma^\mathrm{in}_{k_0} }{\Delta} \approx
%  \frac{8 (\omega_0/\omega_c)^{2/z}}{\Gamma(1+2/z)} %\left(\frac{\omega_0}{\omega_c}\right)^{2/z}
%  \frac{\lambda_1^2}{\Gamma^2} %\left(\frac{\lambda_1}{\Gamma}\right)^2
%  {}_2F_1 \left(1, \frac{1}{z} ,\frac{z+1}{z}, -\frac{4\omega_0^2}{\Gamma^2} %-\left(\frac{\Gamma}{2\omega_0}\right)^2
%  \right),
%\end{align}
\begin{align} \label{eqn:ginel_k0}
%	\negthickspace\negthickspace
	\frac{ 2\pi\Gamma^\mathrm{in}_{k_0} }{\Delta} \approx
	\frac{\pi (\omega_0/\omega_c)^{2/z}}{\Gamma(2/z)\sin(\pi/z)} %\left(\frac{\omega_0}{\omega_c}\right)^{2/z}
\frac{\lambda_1^2}{(\Gamma_0/2)^2} %\left(\frac{\lambda_1}{\Gamma}\right)^2
%	\frac{(\Gamma_0/2)^2}{(\omega_0 - \omega_k)^2 + (\Gamma_0/2)^2}
\left(\frac{\Gamma_0/2}{\omega_0}\right)^{2/z}.
\end{align}
The charge dispersion $\lambda_0$ decreases fast with $\omega_0$, masking the corresponding increase in the number of possible decay channels contributing on-resonance [Fig.~\ref{fig:inelastic_rate}(a)]; this serves to distinguish this process from parasitic effects, such as dielectric loss, which display the opposite behavior~\cite{kuzmin19a}.
We further see that the inelastic scattering probability can approach order unity in the recently-achieved regime of effective fine-structure constant $z \gtrsim 1$~\cite{kuzmin19a,kuzmin19b}. %\cite{devoret07,kuzmin19a,kuzmin19b}.
%
%and ${}_2F_1$ is the hypergeometric function~\cite{abramowitz}.
%Hence, for large $z$, $\Gamma^\mathrm{in}_{\omega_k=\omega_0}/\Delta \sim (\lambda_1/\Gamma)^2 (\Gamma/\omega_0)^{2/z}$. An extension to finite $T \ll \omega_0$ is straightforward.
The increase in number of channels with frequency is seen in an asymmetry of the inelastic resonance lineshape [Fig.~\ref{fig:inelastic_rate}(c,d)]; For $|\omega_k - \omega_0| \ll \omega_0$ one has
\begin{align} \label{eqn:ginel_resonance}
	\negthickspace \frac{ \Gamma^\mathrm{in}_{k} }{\Gamma^\mathrm{in}_{k_0}} \approx
	\begin{cases}
		2\sin \left( \frac{\pi}{z} \right) \left(\frac{\Gamma_0/2}{\omega_k - \omega_0}\right)^{3 - 2/z}, & \omega_k-\omega_0 \gg \Gamma_0, \\ %\omega_0 \gg \omega_k-\omega_0 \gg \Gamma_0 \\
		\frac{(\Gamma_0/2)^2}{(\omega_0 - \omega_k)^2 + (\Gamma_0/2)^2}, & |\omega_k-\omega_0| \lesssim \Gamma_0, \\
		\frac{1-2/z}{\cos ( \pi/z )} \left(\frac{\Gamma_0/2}{\omega_0 - \omega_k}\right)^{4 - 2/z}, & \omega_0 - \omega_k \gg \Gamma_0. %\omega_0 \gg \omega_0 - \omega_k \gg \Gamma_0
	\end{cases}\negthickspace
\end{align}
Finally, let us note that  temperature suppresses coherent quantum phase slips (particularly for $z>1$, when they are relevant~\cite{schon90}) but gives rise to scattering by thermal photons, and hence could either decrease or increase the decay rate, depending on $z$, as shown in Fig.~\ref{fig:inelastic_rate}(e).
Similar expressions to Eqs.~\eqref{eqn:ginel_kkp_knk0}--\eqref{eqn:ginel_resonance} can be obtained via an effective Hamiltonian tailored to describe this particular class of processes~\cite{houzet20}, though that approach cannot give the value of $\omega_c$. %The quality of this approximation is tested in Fig.~\ref{fig:inelastic_rate}(f).

\emph{Quartic nonlinearity.---}
Let us now briefly discuss inelastic photon scattering by more mundane nonlinearities, coming from the Taylor expansion of the transmon Josephson cosine. %, restricting ourselves to $T=0$.
To leading order in $\sqrt{E_C/E_J}$ it is dominated by the Fermi golden rule contribution of the quartic term in the expansion, which at $T=0$ allows an incoming photon at $k$ to split into three at $k_i$, $i=1,2,3$. Expressing $\phi_0$  in terms of the array modes, one finds~\cite[][Sec.~SIII]{SM} %easily finds %a straightforward calculation gives
%\begin{align}
%  \Gamma^\mathrm{in}_{k_1,k_2,k_3|k} = 
%  \frac{\pi}{3} E_C^2 \sin^2(\delta_k) \Pi_{i=1}^3 \sin^2(\delta_{k_i}) %\sin^2(\delta_{k_2}) \sin^2(\delta_{k_3}) 
%  \delta(\omega_k - \omega_{k_1} - \omega_{k_2} - \omega_{k_3}).
%\end{align}
\begin{align} \label{eqn:ginel_quartic}
  \Gamma^\mathrm{in}_{k} = \frac{4 z^2}{3 \pi}
  & \frac{ \omega_0^4 \Delta^4}{\Gamma_0^2}
  \frac{\sin^2 (\delta_{k}) }{\omega_{k}}
  \sum_{k_i} %{k_1,k_2,k_3}
  \frac{\sin^2 (\delta_{k_1}) }{\omega_{k_1}} 
  \times \\ \nonumber & \quad
  \frac{\sin^2 (\delta_{k_2}) }{\omega_{k_2}}
  \frac{\sin^2 (\delta_{k_3}) }{\omega_{k_3}}
  %\prod_{\tilde{k}=k,k_1,k_2,k_3} \frac{\sin^2 (\delta_{\tilde{k}}) }{\omega_{\tilde{k}}}
\delta(\omega_k - \omega_{k_1} - \omega_{k_2} - \omega_{k_3}).
\end{align}
As opposed to the instanton contribution, where $f_k^2$ increases toward low energies [Eq.~\eqref{eqn:fk_inst}], here the factors $\sin^2(\delta_{k_i})/\omega_{k_i} \propto  \omega_{k_i}$ [cf.~Eq.~\eqref{eqn:dk}] suppress the contribution of low frequency photons, and severely limit the phase space for splitting of nearly-resonant photons. %[cf.~Eq.~\eqref{eqn:dk}].
Summing over $k_i$ we find the resulting total inelastic rate near resonance %($\omega_k\approx\omega_0$)
to scale as $\sim z^2 \Delta \Gamma_0^4 / \omega_0^4$. The suppression with $\Gamma_0/\omega_0$ can make it significantly smaller than the instanton contribution, provided $\lambda_1$ is not too small [cf.~Eq.~\eqref{eqn:ginel_k0}]. The ratio between the corresponding rates is depicted in Fig.~\ref{fig:inelastic_rate}(f), which shows that instanton processes are stronger by several orders of magnitude in the experimentally-accessible regime of not-too-small $\Gamma_0/\omega_0$ and $z \gtrsim 1$ [where the exponential factor in Eq.~\eqref{eqn:l0} does not dominate]~\cite{kuzmin19b}.

\emph{Conclusions.---}
In this work we have developed a general formalism for the study of instanton-particle collisions, and applied it to a recently-realized~\cite{kuzmin19b} superconducting circuit in which a transmon qubit is strongly-coupled to a high impedance transmission line. We have shown that significant inelastic single-photon scattering by instantons can be controllably initiated and identified in such a setup: As opposed to the Josephson quartic nonlinearity, which only affects near-resonance photons and thus cannot split them into low frequency ones, an instanton shifts the phases along the entire array, hence couples well to low-$k$ modes, and allows a near-resonant incoming photon to dissipate energy into them.  An experiment has now appeared~\cite{kuzmin20} demonstrating this effect, with favorable comparison to a simplified version of our theory~\cite[][Sec.~SII]{SM}.
This paves the way toward the study of similar effects not only in %high-impedance
various superconducting circuits~\cite{hriscu11,rastelli13,ergul13,vogt16,bard17,schon90,fazio01,camalet04,garciaripoll08,lehur12,goldstein13,peropadre13,snyman15,gheeraert17,leppakangas18,gheeraert18,yamamoto19,belyansky20,forndiaz17,magazzu18,puertasmartinez19,indrajeet20}, but also in other condensed matter (e.g., atomtronic setups)~\cite{leggett,kane92,krinner15} and particle physics~\cite{coleman,gross81,polyakov} systems.

%\begin{acknowledgments}
We thank %R.~Kuzmin and V.~Manucharayn for ongoing collaboration on quantum simulation in circuit QED, and
L.I.~Glazman and M.~Houzet for very useful discussions and for sharing with us their unpublished results~\cite{houzet20}.
Our work has been supported by the U.S.-Israel Binational Science Foundation (Grants No. 2014262 and No. 2016224). In addition, R.V.K.\ and V.E.M.\ acknowledge support from US Department of Energy (Grant No. DE-SC0020160). V.E.M.\ was further supported by a U.S.\ DOE Early Career Award and M.G.\ by the Israel Science Foundation (Grant No. 227/15).
%Our work has been supported by the U.S.-Israel Binational Science Foundation (Grants No.~2014262 and~2016224). V.E.M.\ was further supported by a US DOE Early Career Award, and M.G.\ by the Israel Science Foundation (Grant No.~227/15).
%\end{acknowledgments}
\vspace{-0.5cm}
\bibliography{large_EJ_paper}
\vspace{-50cm}
\include{large_EJ_paper_SM_arxiv}

\end{document}

%% file: large_EJ_paper_SM_arxiv.tex
%\documentclass[aps,pra,singlecolumn,superscriptaddress,floatfix,letterpaper,amssymb]{revtex4-2}
%\documentclass[letterpaper,english,reprint, prl]{revtex4-2}
%%\usepackage[T1]{fontenc}
%%\usepackage[latin9]{inputenc}
%%\setcounter{secnumdepth}{3}
%\usepackage{verbatim}
%\usepackage{amsmath}
%%\usepackage{amssymb}
%\usepackage{graphicx}
%\usepackage{hyperref}
%\usepackage{multirow}
%\usepackage{braket}
%
%\usepackage{bibentry}
%
%\usepackage{color}
%
%%\usepackage{xr-hyper}
%%\externaldocument[MT-]{large_EJ_paper}

%\makeatletter
%
%%%%%%%%%%%%%%%%%%%%%%%%%%%%%%% LyX specific LaTeX commands.
%\pdfpageheight\paperheight
%\pdfpagewidth\paperwidth
%
%%% Because html converters don't know tabularnewline
%\providecommand{\tabularnewline}{\\}
%
%\makeatother
%
%\usepackage{babel}
%\makeatletter
%\input{large_EJ_paper.aux}
%\makeatother

%\newlabel{eqn:fk_inst}{{8}{3}{}{}{}}
%\newlabel{eqn:ginel2_inst}{{9}{3}{}{}{}}
%\newlabel{eqn:pi_inst}{{10}{3}{}{}{}}
%\newlabel{eqn:ginel_kkp}{{11}{3}{}{}{}}

%\begin{document}

\setcounter{equation}{0}
\setcounter{figure}{0}
\setcounter{table}{0}
%\setcounter{page}{1}
%\makeatletter
\renewcommand{\theequation}{S\arabic{equation}}
\renewcommand{\thefigure}{SF\arabic{figure}}
\renewcommand{\thetable}{ST\arabic{table}}
\renewcommand{\thesection}{S\Roman{section}}

\begin{widetext}

\section*{Supplemental Material for: ``Photon-instanton collider implemented by a superconducting circuit''}

%\author{Amir Burshtein}
%\affiliation{Raymond and Beverly Sackler School of Physics and Astronomy, Tel Aviv University, Tel Aviv 6997801, Israel}
%
%\author{Roman Kuzmin}
%\affiliation{Department of Physics, University of Maryland, College Park, MD 20742, USA}
%%\affiliation{Department of Physics, Joint Quantum Institute, and Center for Nanophysics and Advanced Materials, University of Maryland, College Park, MD 20742, USA}
%
%\author{Vladimir E. Manucharyan}
%\affiliation{Department of Physics, University of Maryland, College Park, MD 20742, USA}
%%\affiliation{Department of Physics, Joint Quantum Institute, and Center for Nanophysics and Advanced Materials, University of Maryland, College Park, MD 20742, USA}
%
%\author{Moshe Goldstein}
%\affiliation{Raymond and Beverly Sackler School of Physics and Astronomy, Tel Aviv University, Tel Aviv 6997801, Israel}
%
%%\date{\today}
%\begin{abstract}
In this Supplemental Material we provide additional technical details and elaborate upon some of the derivations in the main text.
In Sec.~\ref{sec:instanton}, we provide some intermediate steps in the derivation of the instanton inelastic rate which were omitted from the main text.
In Sec.~\ref{sec:dual_cosine}, we use a phenomenological ``dual-cosine'' approach to calculate the rate, and compare its results with those of the instanton approach.
Finally, in Sec.~\ref{sec:quartic}, we derive the inelastic rate due to the Josephson cosine quartic nonlinearity. %, and provide a closed-form formula for a photon that is close to resonance.

%\end{abstract}
%\maketitle

\section{Additional details of the derivation of the instanton inelastic rate}
\label{sec:instanton}
The main result of the paper is the inelastic rate due to the photon-instanton interaction, Eqs.~\eqref{eqn:pi_inst}-\eqref{eqn:ginel_kkp} of the main text, which apply to incoming/outgoing photons with frequencies between $\lambda_*$ and $3\omega_0$, assuming in addition that $\lambda_* \ll \max(\Gamma_0,T) \ll \omega_0$ and $E_C \ll E_J$ (all these parameters will be redefined below).
In this section, we repeat the derivation while providing some additional details.
%, and fill in the gaps between the steps pointed out in the main text.
We recite the Lagrangian of the system, Eq.~\eqref{eqn:lagrangian} [describing the simplified circuit depicted in Fig.~\ref{fig:circuit}(b)] of the main text:
\begin{equation} \label{eqnS:lagrangian}
\negthickspace
\mathcal{L} = \frac{C_0 \dot{\phi}_0^2}{2} + E_J \cos \left( 2\phi_0 \right) + \sum_{n=1}^{N} %\left[
\frac{C_g \dot{\phi}_n^2}{2} - \frac{\left(\phi_n - \phi_{n-1} \right)^2}{2L}, %\right],
\end{equation}
where $\phi_n$ is the flux of node $n = 0 \cdots N$ in Fig.~\ref{fig:circuit}(b), $L$ and $C_g$ are the array effective Josephson inductance and ground capacitance, respectively, $E_J$ and $C_0$ are the transmon Josephson energy and capacitance, respectively, and
%As in the main text,
we employ units where $e=1$ and $\hbar=1$. %In the following, we show how one may derive the inelastic rate from this Lagrangian.

\subsection{An isolated transmon}
The Hamiltonian of an isolated Josephson junction~\cite{koch07} corresponds to the first two terms of the Lagrangian, Eq.~\eqref{eqnS:lagrangian}:
\begin{equation} \label{eqnS:trans_ham}
H_0 = E_C (Q_0 - q_g)^2 - E_J \cos(2\phi_0),
\end{equation}
where  $E_C=1/2C_0$, $\phi_0$ and $Q_0$ are the flux and charge variables of the transmon, obeying the canonical commutation relations, and $q_g$ is its gate charge (which can be gauged away in the presence of an infinite transmission line, hence is ignored in later sections).
This Hamiltonian may be solved exactly in the flux basis, where the energy of the $m$th level, or Bloch band, $E_m(q_g) = E_m(q_g \pm 2)$ ($m=0,1,\cdots$), is given by the Mathieu characteristic values $a_\nu(E_J/2E_C)$ and $b_\nu(E_J/2E_C)$~\cite{abramowitz}, as summarized in Table~\ref{tab:mathieu}.
The transmon resonance frequency $\omega_0$ is defined by
\begin{equation} \label{eqnS:omega_01_mathieu}
\omega_{0}(q_g) = E_1(q_g) - E_0(q_g),
\end{equation}
and generally depends on $q_g$. In the transmon limit, $E_J \gg E_C$, the effect of charge offset is negligible, and $\omega_0 \approx \sqrt{8 E_J E_C}$ to leading order in $E_C/E_J$, which is also the resonance frequency which results from approximating the Josephson cosine potential in Eq.~\eqref{eqnS:trans_ham} by a harmonic term, $2 E_J \phi_0^2$. The lowest order correction may be calculated in perturbation theory due to the quartic term in the expansion of $\cos(2\phi_0)$, and yields $\omega_0 \approx \sqrt{8 E_J E_C} - E_C$.

%As mentioned in the main text, $\lambda_0$ is given by the half bandwidth of the lowest Bloch band of the Mathieu equation:
%\begin{equation} \label{eqnS:lambda_0}
%\frac{\lambda_0}{E_C} = \frac{E_0(n_g = 1/2) - E_0(n_g = 0)}{2} = \frac{b_1(q) - a_0(q)}{2} \approx \frac{8}{\sqrt{\pi}} \left( 2 E_J/E_C \right)^{3/4} e^{-\sqrt{8E_J/E_C}},
%\end{equation}
%where $a_0(q)$ and $b_1(q)$ are the Mathieu eigenvalues \textcolor{blue}{ABRAMOWITZ REFERENCE}. The approximated form is the WKB approximation (given also by Eq.~\eqref{eqn:l0} of the main text) which results from the instanton trajectory, $\phi_0^{(0)}(\tau) = \pm 2\tan^{-1}(e^{\omega_0 \tau})$. Another useful quantity is the charge dispersion of the first excited state and its WKB approximation:
%\begin{equation} \label{eqnS:lambda_1}
%\frac{\lambda_1}{E_C} = \frac{E_1(n_g = 1/2) - E_1(n_g = 0)}{2} = \frac{b_2(q) - a_1(q)}{2} \approx -\frac{8}{\sqrt{\pi}}\sqrt{128 E_J/E_C} \left(2 E_J / E_C\right)^{3/4} e^{-\sqrt{8 E_J / E_C}}.
%\end{equation}
%In Eq.~(4) of the main text, we use $C_k$, the capacitance of mode $k$.
We define the charge dispersion $\lambda_m$ as the half-bandwidth of the $m$th Bloch band, given by
\begin{equation} \label{eqnS:lambda_m}
\frac{\lambda_m}{E_C} = \frac{E_m(q_g = 1) - E_m(q_g = 0)}{2 E_C} = \frac{(-1)^m}{2} \left[ b_{m+1}\left(\frac{E_J}{2E_C}\right) - a_m\left(\frac{E_J}{2E_C}\right) \right] \approx \frac{2^{4m + 9/2} (-1)^m}{m!\sqrt{\pi}} \left(\frac{E_J}{2E_C}\right)^{m/2 + 3/4} e^{-\sqrt{8E_J/E_C}},
\end{equation}
%where $a_m(q)$ and $b_{m+1}(q)$ are the Mathieu eigenvalues \textcolor{blue}{ABRAMOWITZ REFERENCE}.
The final form is the result of the WKB approximation [given also by Eq.~\eqref{eqn:l0} of the main text for $m=0$]. In the imaginary-time path integral formalism it can be derived from the action of and quadratic fluctuations around the instanton solution, $\phi_0^{(0)}(\tau) = \pm 2\tan^{-1}(e^{\omega_0 \tau})$, which interpolates between two adjacent minima of the cosine~\cite{marino,schon90,fazio01}. Its Fourier transform is $\phi_0^{(0)} (\omega) = \pm \pi / i\omega \cosh \left( \pi \omega/2 \omega_0 \right)$. Here and below, the plus and minus signs correspond to an instanton and an anti-instanton, respectively.

\begin{table}
	\begin{ruledtabular}
		\begin{tabular}{c|ccccc}
			$E_m(q_g)/E_C$ & $q_g=0$         &  & $0<|q_g|<1$                                    & & $|q_g|=1$ \\
			\hline
			$m$ even       & $a_m(E_J/2E_C)$ & $<$ & $a_{m+|q_g|}(E_J/2E_C) = b_{m+|q_g|}(E_J/2E_C)$ & $<$ & $b_{m+1}(E_J/2E_C)$ \\
			$m$ odd  & $b_{m+1}(E_J/2E_C)$ & $>$ & $a_{m+1-|q_g|}(E_J/2E_C) = b_{m+1-|q_g|}(E_J/2E_C)$ & $>$ & $a_m(E_J/2E_C)$
			\\
		\end{tabular}
	\end{ruledtabular}
	\caption{\label{tab:mathieu} The energy levels of an isolated transmon $E_m$ ($m=0,1,\cdots$) as function of the quasicharge $q_g$ in the first Brillouin zone $-1<q_g<1$ in terms of the Mathieu characteristic values $a_\nu$, $b_\nu$~\cite{abramowitz}.}
\end{table}

\subsection{The array modes and their effect on the instanton action}
In the presence of the array, a phase slip in the transmon perturbs the array modes; the corresponding leading contribution to the action is given by Eq.~\eqref{eqn:fk_tilde} of the main text. Here we provide its derivation. %some details regarding the derivation of $\delta S$.
To set the stage, let us first consider the limit where $E_J/E_C$ is large enough for the Josephson cosine in Eq.~\eqref{eqnS:lagrangian} to be approximated by a harmonic potential, so that
%We begin by writing down the harmonic version of the Lagrangian given by Eq.~\eqref{eqnS:lagrangian}:
\begin{equation} \label{eqnS:lag_harm}
\mathcal{L} \approx \frac{C_0 \dot{\phi}_0^2}{2} - 2 E_J \phi_0^2 + \sum_{n=1}^{N}
\frac{C_g \dot{\phi}_n^2}{2} - \frac{\left(\phi_n - \phi_{n-1} \right)^2}{2L},
\end{equation}
and the equations of motion are given by
\begin{align} \label{eqnS:eqs_of_motion}
-C_g \ddot{\phi}_n &= \frac{1}{L}\left(2\phi_n - \phi_{n - 1} - \phi_{n + 1} \right), & n \ge 1, \nonumber \\
-C_0 \ddot{\phi}_0 &= 4 E_J \phi_0 + \frac{1}{L}\left(\phi_0 - \phi_1 \right),  & n = 0.
\end{align}
The solutions are of the form $\phi_n \propto \sin(kn + \delta_k)$. The equations for $n\ge 1$ yield the dispersion relation $\omega_k = 2v\sin(k/2) \approx v k$ (the approximation holds for $k \ll \pi$, that is, $\omega_k \ll v$), with $v = 1/\sqrt{LC_g}$. The mode spacing is $\Delta = \pi v/N$ (hence $\sum_k \to \int_0^\infty\mathrm{d}\omega/\Delta$), while the wave impedance is $Z=\sqrt{L/C_g} \equiv  z\pi/2$. The equation for $n=0$ yields the phase shift:
\begin{equation} \label{eqnS:dk}
\tan\delta_k = \frac{\Gamma_0 \omega_k \sqrt{1 - \left(\frac{\omega_k}{2v}\right)^2}}{\omega_0^2 - \omega_k^2 \left(1 - \frac{\Gamma_0}{2v}\right)} \approx \frac{\Gamma_0 \omega_k}{\omega_0^2 - \omega_k^2},
\end{equation}
where $\Gamma_0 = 1/Z C_0 = 4 E_C / \pi z$ is the elastic resonance broadening.
The approximate form corresponds to Eq.~\eqref{eqn:dk} of the main text. The $1 - \Gamma_0/2v$ factor in the denominator of the exact form only gives a slight shift of the resonance frequency, and may always be replaced by unity. The $\sqrt{1 - (\omega_k/2v)^2}$ factor in the numerator, however, should be retained when modes with $k \sim \pi$ contribute, as we discuss below.
%$\tan\delta_k = (\Gamma_0 \omega_k)/(\omega_0^2 - \omega_k^2)$, also given by Eq.~\eqref{eqn:dk} of the main text.
%The $k$ mode capacitance, $C_k = NC_g/2$, is found using the orthogonality relations of Eq.~\eqref{eqnS:lag_harm}.

%The action due to an instanton $\phi_0^{(0)} = \pm 2\tan^{-1}(e^{\omega_0 \tau})$ is given by Eq.~\eqref{eqn:ds} of the main text. The leading terms are:
Let us now return to $E_J \gtrsim E_C$, where instantons may appear. For $\omega_0/\Gamma_0 \gg 1$ we may expand the imaginary time dynamics around the classical isolated instanton solution, $\phi_0^{(0)}(\tau) = \pm 2\tan^{-1}(e^{\omega_0 \tau})$ and $\phi_{n > 0}^{(0)}(\tau)=0$.
%An instanton occuring at the transmon would induce flux deviations throughout the array.
We thus set $\phi_n(\tau) = \phi_n^{(0)}(\tau) + \delta\phi_n(\tau)$, %, with $\phi_0^{(0)}(\tau)$ given by the instanton trajectory mentioned above and $\phi_{n > 0}^{(0)}(\tau)=0$.
%Expanding the deviations using the $k$ modes,
and expand $\delta\phi_n = \sum_k \delta\phi_k \sin(k n+\delta_k)$. To second order in $\delta\phi_k$ we obtain the following action [Eq.~\eqref{eqn:ds} of the main text], 
\begin{align} \label{eqnS:ds_1}
S = S_0 & + \int \frac{\mathrm{d} \omega}{2\pi}
\left[
\frac{\left|\phi_0^{(0)} (\omega)\right|^2}{2L}
+  \sum_k %\left[
\frac{C_k}{2} (\omega^2+\omega_k^2) \left| \delta\phi_k(\omega) \right|^2 - \frac{\sin(k+\delta_k) - \sin(\delta_k)}{L} \phi_0^{(0)} (-\omega) \delta \phi_k (\omega) \right] \nonumber \\
& - \int \mathrm{d}\tau %\left\{%\sum_{k,k^\prime}
\frac{8 E_J}{\cosh^2 (\omega_0\tau)} \left[ \sum_k \sin(\delta_k)  \delta \phi_k(\tau) \right]^2,
\end{align}
where $S_0 = \sqrt{8E_J/E_C}$ is the classical action of an isolated instanton and $C_k \approx N C_g/2$ is the capacitance of mode $k$. The second row of Eq.~\eqref{eqnS:ds_1} yields subleading terms in $\Gamma_0/\omega_0$ (as can be verified by plugging into it the solution we obtain below) and will be neglected henceforth.
%\begin{align} \label{eqn:ds}
%S = S_0 & + \int \frac{\mathrm{d} \omega}{2\pi}
%\left[
%\frac{\left|\phi_0^{(0)} (\omega)\right|^2}{2L}
%+  \sum_k %\left[
%\frac{C_k}{2} (\omega^2+\omega_k^2) \left| \delta\phi_k(\omega) \right|^2
%\right.%\right.
%\nonumber \\
%& \qquad \quad
%\left.%\left.
%- \frac{\sin(k+\delta_k) - \sin(\delta_k)}{L} \phi_0^{(0)} (-\omega) \delta \phi_k (\omega) \right]
%%\right\}
%\nonumber \\
%& %\negthickspace %\negthickspace %\negthickspace %\negthickspace
%- \int \mathrm{d}\tau %\left\{%\sum_{k,k^\prime}
%\frac{8 E_J}{\cosh^2 (\omega_0\tau)} \left[ \sum_k \sin(\delta_k)  \delta \phi_k(\tau) \right]^2, %\right\}, %\sin(\delta_{k^\prime}) \delta \phi_{k^\prime}(\tau).
%\end{align}
Solving the equation of motion for $\delta\phi_k(\omega)$ and using the $n=0$ part of Eq.~\eqref{eqnS:eqs_of_motion}, we find
\begin{align} \label{eqnS:dphik}
\delta \phi_k(\omega) = \frac{2v^2}{N} \frac{\sin(k + \delta_k) - \sin(\delta_k)}{\omega^2 + \omega_k^2} \phi_0^{(0)} (\omega) = \frac{\sin(\delta_k)}{Z C_k \Gamma_0} \frac{\omega_0^2 - \omega_k^2}{\omega^2 + \omega_k^2} \phi_0^{(0)} (\omega) = \frac{\omega_k\cos(\delta_k)\sqrt{1 - \left(\frac{\omega_k}{2v}\right)^2}}{Z C_k} \frac{\phi_0^{(0)} (\omega)}{\omega^2 + \omega_k^2},
\end{align}
%\begin{align} \label{eqnS:dphik}
%\delta \phi_k(\omega) &= \frac{2v^2}{N} \frac{\sin(k + \delta_k) - \sin(\delta_k)}{\omega^2 + \omega_k^2} \phi_0^{(0)} (\omega) = \frac{\sin(\delta_k)}{Z C_k \Gamma_0} \frac{\omega_0^2 - \omega_k^2}{\omega^2 + \omega_k^2} \phi_0^{(0)} (\omega) \nonumber \\
%&= \frac{\omega_0^2 - \omega_k^2}{\sqrt{\left(\omega_0^2 - \omega_k^2\right)^2\left(1 - \frac{\omega_k}{2v}\right)^2 + (\Gamma_0\omega_k)^2}}\frac{\omega_k \sqrt{1 - \left(\frac{\omega_k}{2v}\right)^2}}{Z C_k} \frac{\phi_0^{(0)} (\omega)}{\omega^2 + \omega_k^2},
%\end{align}
in correspondence with Eq.~\eqref{eqn:dphik} of the main text, but with the additional factor $\sqrt{1 - (\omega_k/2v)^2}$ which is approximately unity for $\omega_k \ll v$. Plugging this back to Eq.~\eqref{eqnS:ds_1}, the correction to the action $\delta S = S - S_0$ becomes
%\begin{align} \label{eqnS:ds_2}
%\delta S = \int \frac{\mathrm{d} \omega}{2\pi} \frac{\left|\phi_0^{(0)} (\omega)\right|^2}{2L}
%\left[1
%- \frac{2}{N}\sum_k %\left[
%\frac{\left(\omega_0^2 - \omega_k^2\right)^2 \sin^2(\delta_k)}{\left(\omega_k\Gamma_0\right)^2} + \frac{2}{N}\sum_k %\left[
%\frac{\left(\omega_0^2 - \omega_k^2\right)^2 \omega^2 \sin^2(\delta_k)}{\left(\omega_k\Gamma_0\right)^2(\omega^2+\omega_k^2)} \right].
%\end{align}
\begin{align} \label{eqnS:ds_2}
\delta S = \int \frac{\mathrm{d} \omega}{2\pi} \frac{\left|\phi_0^{(0)} (\omega)\right|^2}{2L}
\left[1
- \frac{2}{N}\sum_k %\left[
\left(1 - (\omega_k/2v)^2\right)\cos^2(\delta_k) + \frac{2}{N}\sum_k %\left[
\frac{\omega^2 \left(1 - (\omega_k/2v)^2\right)\cos^2(\delta_k)}{\omega^2+\omega_k^2} \right].
\end{align}
We note that in both of the sums over $k$ in Eq.~\eqref{eqnS:ds_2} we may approximate $\cos^2(\delta_k) \approx 1$, since the deviation of the cosine from $1$ occurs only in a frequency range of order $\Gamma_0$ around the resonance, and hence contributes subleading terms in $\Gamma_0/\omega_0$. Care must be taken in the evaluation of the first sum over $k$ in Eq.~\eqref{eqnS:ds_2}, since the argument does not decay for $k \sim \pi$, and therefore we must account for terms of order $\omega_k/2v$, including the Jacobian $\mathrm{d}\omega_k / \mathrm{d}k = \sqrt{1 - (\omega_k/2v)^2}$. This is not the case for the second sum, where the summand decays rapidly outside the regime $k \ll \pi$, and therefore we may drop all terms of order $\omega_k/v$. Overall, we find:
\begin{equation} \label{eqnS:delta_S}
\delta S = \frac{1}{2 z} \int_{-\infty}^{\infty} \frac{\mathrm{d}\omega}{|\omega|\cosh^2\left(\frac{\pi}{2}\frac{\omega}{\omega_0}\right)} = \frac{\Delta}{z} \sum_k \frac{1}{\omega_k \cosh^2\left(\frac{\pi}{2}\frac{\omega_k}{\omega_0}\right)} \equiv
\frac{1}{2} \sum_k \tilde{f}_k^2,
\end{equation}
where %we used the symmetry of the integrand and expressed the $\omega$ integral as a sum over the modes $k$ using the density of states.
we have rewritten the $\omega$ integral as a $k$ sum.
We thus arrive at Eq.~\eqref{eqn:fk_tilde} of the main text.
Combining $\delta S$ with $S_0$, and noting that to leading order in $\Gamma_0/\omega_0$ Gaussian fluctuations around the classical solution we have just discussed can be approximated by their value for an isolated transmon, the overall amplitude of a single instanton is $(\lambda_0/2) e^{-\sum_k \tilde{f}_k^2/2}$.
As discussed in the main text, for $z>1$ instantons are a relevant perturbation, giving rise to an emergent scale, $\lambda_* \sim \lambda_0 (\lambda_0/\omega_0)^{1/(z-1)}$~\cite{schon90}. Below $\lambda^*$ instanton effects are nonperturbative; in the following we concentrate on higher energies.
%\begin{equation}
%\delta S = \frac{1}{4 v L}\int_{-\infty}^{\infty} \mathrm{d}\omega \frac{1}{\cosh^2\left(\frac{\pi \omega}{2 \omega_0}\right)} \int_{0}^{\infty} \frac{\mathrm{d}\omega_k}{\omega^2 + \omega_k^2} = \frac{1}{8 z}\int_{-\infty}^{\infty} \mathrm{d}\omega \frac{1}{|\omega| \cosh^2\left(\frac{\pi \omega}{2 \omega_0}\right)}
%\end{equation}

\subsection{Calculation of the $\mathcal{T}$-matrix}
To calculate the inelastic rate, we need the elements of the $\mathcal{T}$-matrix corresponding to all possible inelastic processes. %The inelastic rate is calculated using the Fermi golden rule for he rate each possible processes, and summing over these processes. $\mathcal{T}$-matrix elements.
According to the imaginary-frequency version of the Lehmann-Symanzik-Zimmermann (LSZ) reduction formula~\cite{peskin,zhou96}, these  matrix elements for multiple incoming and outgoing photons are determined by the multipoint correlation function of the involved photons, with the single-particle legs (Green functions) amputated:
\begin{align} \label{eqnS:lsz}
\mathcal{T}^{ k^\prime_1, k^\prime_2, \cdots, k^\prime_{N_\mathrm{in}} }_{ k_1, k_2, \cdots, k_{N_\mathrm{out}} } & =
\frac{\Delta}{2\pi}
\frac{G_0^{-1}\left(\omega^\prime_1, k^\prime_1\right)}{ \sqrt{2 C_{k^\prime_1} \omega_{k^\prime_1} } }
\cdots
\frac{G_0^{-1}\left(\omega^\prime_{N_\mathrm{in}}, k^\prime_{N_\mathrm{in}}\right)}{ \sqrt{2 C_{k^\prime_{N_\mathrm{in}}} \omega_{k^\prime_{N_\mathrm{in}}} } }
\frac{G_0^{-1}\left(\omega_1, k_1\right)}{ \sqrt{2 C_{k_1} \omega_{k_1} } }
\cdots
\frac{G_0^{-1}\left(\omega_{N_\mathrm{out}}, k_{N_\mathrm{out}}\right)}{ \sqrt{2 C_{k_{N_\mathrm{out}}} \omega_{k_{N_\mathrm{out}}} } }
\times \nonumber \\ &
\left\langle
\phi_{k^\prime_1}(\omega^\prime_1) \cdots \phi_{k^\prime_{N_\mathrm{in}}} (\omega^\prime_{N_\mathrm{in}})
\phi_{k_1}(\omega_1) \cdots \phi_{k_{N_\mathrm{out}}} (\omega_{N_\mathrm{out}})
\right\rangle_\mathrm{1-instanton}
\Big|
^{ \omega^\prime_1 \to i \omega_{k^\prime_1}, \cdots, \omega^\prime_{N_\mathrm{in}} \to i \omega_{k^\prime_{N_\mathrm{in}}} }
_{ \omega_1 \to -i \omega_{k_1}, \cdots, \omega_{N_\mathrm{out}} \to -i \omega_{k_{N_\mathrm{out}}} },
\end{align}
with $G_0(\omega, k) = 1/C_k(\omega^2 + \omega_k^2)$ being the Green function of mode $k$. The $1/\sqrt{2 C_k \omega_k}$ prefactors relate $\phi_k$ to $a_k^\dagger$ and $a_k$, the creation and annihilation of operators of mode $k$, respectively. We thus recover the first line of Eq.~\eqref{eqn:T} of the main text. The single-instanton propagator may be calculated using the path integral formalism. As mentioned above, each instanton contributes an amplitude $(\lambda_0/2)e^{-\sum_k \tilde{f}_k^2/2}$, which stems from the classical action $S_0 + \delta S$ together with Gaussian fluctuations around the classical path. To leading order in $E_C/E_J$ and $\Gamma_0/\omega_0$ we may replace the fields in the multipoint correlators by their values at the classical instanton path,
%\begin{align}
%\left\langle
%\phi_{k^\prime_1}(\omega^\prime_1) \cdots \phi_{k^\prime_{N_\mathrm{in}}} (\omega^\prime_{N_\mathrm{in}})
%\phi_{k_1}(\omega_1) \cdots \phi_{k_{N_\mathrm{out}}} (\omega_{N_\mathrm{out}})
%\right\rangle_\mathrm{1-instanton}
%\Big|
%^{ \omega^\prime_1 \to i \omega_{k^\prime_1}, \cdots, \omega^\prime_{N_\mathrm{in}} \to i \omega_{k^\prime_{N_\mathrm{in}}} }
%_{ \omega_1 \to -i \omega_{k_1}, \cdots, \omega_{N_\mathrm{out}} \to -i \omega_{k_{N_\mathrm{out}}} } &=
%\nonumber \\
%&\hspace{-4cm}=
%\end{align}
\begin{equation} \label{eqnS:propagator}
\left\langle
\ldots\right\rangle
%\Big|
%^{ \omega^\prime_1 \to i \omega_{k^\prime_1}, \cdots, \omega^\prime_{N_\mathrm{in}} \to i \omega_{k^\prime_{N_\mathrm{in}}} }
%_{ \omega_1 \to -i \omega_{k_1}, \cdots, \omega_{N_\mathrm{out}} \to -i \omega_{k_{N_\mathrm{out}}} }
\Big|
^{ \omega^\prime_1 \to i \omega_{k^\prime_1}, \cdots}
_{ \omega_1 \to -i \omega_{k_1}, \cdots} = (\pm 1)^{N_\mathrm{in} + N_\mathrm{out}} \frac{\lambda_0}{2} e^{-\sum_k \tilde{f}_k^2/2} 
\delta\phi_{k^\prime_1}(\omega^\prime_1) \cdots \delta\phi_{k^\prime_{N_\mathrm{in}}} (\omega^\prime_{N_\mathrm{in}})
\delta\phi_{k_1}(\omega_1) \cdots \delta\phi_{k_{N_\mathrm{out}}} (\omega_{N_\mathrm{out}}),
\end{equation}
where, as mentioned above, the plus and minus signs correspond to instantons and anti-instantons, respectively. Plugging Eq.~\eqref{eqnS:propagator} back into Eq.~\eqref{eqnS:lsz} and taking the limits $\omega_j \to -i \omega_{k_j}, \omega^\prime_j \to i \omega_{k^\prime_j}$, one finds
\begin{equation} \label{eqnS:T_mat}
\mathcal{T}^{ k^\prime_1, k^\prime_2, \cdots, k^\prime_{N_\mathrm{in}} }_{ k_1, k_2, \cdots, k_{N_\mathrm{out}} } = (\mp 1)^{N_\text{in}} (\pm 1)^{N_\text{out}} f_{k^\prime_1} f_{k^\prime_2} \cdots f_{ k^\prime_{N_\mathrm{in}} }
f_{k_1} f_{k_2} \cdots f_{ k_{N_\mathrm{out}} }
\frac{\lambda_0}{2} e^{-\sum_k \tilde{f}_k^2/2},
\end{equation}
with
\begin{equation} \label{eqnS:f_k}
f_k = \sqrt{\frac{2\Delta}{z\omega_k}}
\frac{ \omega_0^2 - \omega_k^2 }{ \sin \left(\frac{\pi}{2}\frac{\omega_0-\omega_k}{\omega_0} \right) \sqrt{\left( \omega_0^2 - \omega_k^2 \right)^2 + (\Gamma_0 \omega_k)^2}},
\end{equation}
which are Eqs.~\eqref{eqn:T}--\eqref{eqn:fk_inst} of the main text.

\subsection{Calculation of the inelastic rate} \label{ssec:ginel}
The $\mathcal{T}$-matrix elements in Eq.~\eqref{eqnS:T_mat} allow us to calculate the inelastic rate using the Fermi golden rule. Before taking the squared absolute value, we have to sum over the contributions of instantons and anti-instantons in Eq.~\eqref{eqnS:T_mat}. We see that the matrix elements vanish for odd $N_\mathrm{out}+N_\mathrm{in}$, and on the other hand are multiplied by 2 for even $N_\mathrm{out}+N_\mathrm{in}$; this reflects the symmetry of the Lagrangian, Eq.~\eqref{eqnS:lagrangian} upon flipping all the phases.
Let us now consider a process in which a photon at mode $k$ collides with photons at $k^\prime_1, k^\prime_2, \cdots, k^\prime_{N_\mathrm{in}}$ (so now there are $N_\mathrm{in}+1$ incoming photons) to produce photons at $k_1, k_2, \cdots, k_{N_\mathrm{out}}$, with odd $N_\mathrm{out}+N_\mathrm{in} \ge 3$; by the Fermi golden rule, the rate of such a process is
%Squaring the $\mathcal{T}$-matrix elements (after summing over the instanton and anti-instanton contributions) and multiplying them by the appropriate Bose-Einstein factors and energy conservation delta functions The rate of a process involving several incoming and outgoing photons with odd $N_\mathrm{out}+N_\mathrm{in} \ge 3$ is thus given by
\begin{equation} \label{eqnS:FGR}
\Gamma^{ k, k^\prime_1, k^\prime_2, \cdots, k^\prime_{N_\mathrm{in}} }_{ k_1, k_2, \cdots, k_{N_\mathrm{out}} } = \lambda_0^2 f_k^2 e^{-\sum_{k^\prime} \tilde{f}_{k^\prime}^2} f_{k_1}^2 \cdots f_{k_{N_\mathrm{out}}}^2 f_{k_1^\prime}^2 \cdots f_{k^\prime_{N_\mathrm{in}}}^2
2\pi \delta \left( \omega_k + \omega_{k^\prime_1} + \cdots + \omega_{k^\prime_{N_\mathrm{in}}} -\omega_{k_1} - \cdots -\omega_{k_{N_\mathrm{out}}} \right).
\end{equation}
%The total inelastic rate is then found by squaring the $\mathcal{T}$-matrix elements (after summing over the instanton and anti-instanton contributions), multiplying them by the appropriate Bose-Einstein factors and energy conservation delta functions, and then summing over the momenta and over all possible number of incoming and outgoing photons, for odd $N_\mathrm{out}+N_\mathrm{in} \ge 3$.
However, given a specific process, we note that we might add photons without changing the outcome. For instance, adding a single photon at mode $k^{\prime\prime}$ to both the incoming and outgoing photons would result in a process that is physically identical to the one without it, but with a $\mathcal{T}$-matrix element multiplied by $-f_{k^{\prime\prime}}^2$ [cf.~Eq.~\eqref{eqnS:T_mat}]. Adding this amplitude to the amplitude excluding the photon at $k^{\prime\prime}$ we thus find that when such a photon is present in the system, the rate, Eq.~\eqref{eqnS:FGR}, would be multiplied by a factor $(1 - f_{k^{\prime\prime}}^2)^2$. Now, at finite temperature $T$ the occupation of this mode is set by the Bose-Einstein distribution, $n_B(\omega_{k^{\prime\prime}})$, leading to a factor
\begin{align}
	1 + \left[ (1 - f_{k^{\prime\prime}}^2)^2 -1 \right] n_B(\omega_{k^{\prime\prime}}) \approx 1 - 2f_{k^{\prime\prime}}^2 n_B(\omega_{k^{\prime\prime}}) \approx e^{-2 f_{k^{\prime\prime}}^2 n_B(\omega_{k^{\prime\prime}})},
\end{align}
where the approximations are based on the smallness of $f_{k^{\prime\prime}}^2 \sim 1/N$ [see Eq.~\eqref{eqnS:f_k}] in the thermodynamic limit (similar considerations relieve us from the need to explicitly treat cases where some of the photon modes in the forthcoming equations coincide).
%To avoid double-counting, we have to explicitly exclude photons that are not present in Eq.~\eqref{eqnS:prob_inst}. Since each photon contributes a factor $f_{k^{\prime\prime}}^2$, the probability of a process that includes a photon $k^{\prime\prime}$ as either an incoming or outgoing photon would be multiplied by $1 - (1 - f_{k^{\prime\prime}}^2)^2 \approx 2 f_{k^{\prime\prime}}^2$, where we drop a factor $f_{k^{\prime\prime}}^4 \sim 1/N^2$ in the thermodynamic limit. Accounting for the Bose-Einstein factor of $k^{\prime\prime}$, we see that $\mathcal{P}$ should be multiplied by factors $1 - 2f_{k^{\prime\prime}}^2 n_B(\omega_{k^{\prime\prime}})$ for all $k^{\prime\prime}$ modes that are not involved in the process. The inelastic rate is then given by
Multiplying Eq.~\eqref{eqnS:FGR} by this factor for all possible $k^{\prime\prime}$, as well as by the appropriate Bose-Einstein factors for the photons at $k^\prime_1, k^\prime_2, \cdots, k^\prime_{N_\mathrm{in}}$ and $k_1, k_2, \cdots, k_{N_\mathrm{out}}$, and subtracting the contribution of processes in which the photon at $k$ is emitted rather than absorbed, we recover Eq.~\eqref{eqn:ginel2_inst} of the main text:
\begin{align} \label{eqnS:ginel2_inst_1}
	\Gamma^\mathrm{in}_k = &
	\lambda_0^2 f_k^2 e^{-\sum_{k^\prime} \tilde{f}_{k^\prime}^2  - 2 \sum_{k^\prime} f_{k^\prime}^2 n_B(\omega_{k^\prime})}
	\sum_{N_\mathrm{out},N_\mathrm{in}}
	\sum_{\substack{k_1<\cdots<k_{N_\mathrm{out}},\\ k_1^\prime<\cdots<k^\prime_{N_\mathrm{in}}}}
	f_{k_1}^2 \cdots f_{k_{N_\mathrm{out}}}^2 f_{k_1^\prime}^2 \cdots f_{k^\prime_{N_\mathrm{in}}}^2
	(1+n_B (\omega_{k_1})) \cdots (1+n_B (\omega_{k_{N_\mathrm{out}}}))
	\times \nonumber
	\\ %\nonumber
	&	
	%\negthickspace\negthickspace\negthickspace\negthickspace
	%(1+n_B (\omega_{k_1})) \cdots (1+n_B (\omega_{k_{N_\mathrm{out}}}))
	\qquad\qquad
	n_B (\omega_{k^\prime_{1}}) \cdots n_B (\omega_{k^\prime_{N_\mathrm{in}}}) 
	2\pi \left[ \delta \left( \omega_k + \omega_{k^\prime_1} + \cdots + \omega_{k^\prime_{N_\mathrm{in}}} -\omega_{k_1} - \cdots -\omega_{k_{N_\mathrm{out}}} \right)  - \{\omega_k \to -\omega_k\} \right],
\end{align}
where the sum is over odd $N_\mathrm{out}+N_\mathrm{in} \ge 3$.

We now move on to express $\Gamma_k^{\mathrm{in}}$ in terms of the retarded self energy $\Pi_R(\omega)$, given by Eq.~\eqref{eqn:pi_inst} of the main text. First, we express the delta functions in terms of their Fourier representation, $2\pi\delta(\omega) = 2\Re\left\{\int_{0}^{\infty}\mathrm{d}te^{i\omega t}\right\}$. %, which yields a product of time-dependent exponentials.
%We then assign each form factor its corresponding exponential factor, and push the
Then we rearrange the sums over the momenta into a product (lifting the restrictions $k_1<\cdots<k_{N_\mathrm{out}}$, $k_1^\prime<\cdots<k^\prime_{N_\mathrm{in}}$ gives a factor $1/N_\mathrm{out}! N_\mathrm{in}!$):
%\begin{align} \label{eqnS:ginel2_inst_1}
%\Gamma^\mathrm{in}_k = &
%\lambda_0^2 f_k^2 e^{-\sum_{k^\prime} \tilde{f}_{k^\prime}^2  - 2 \sum_{k^\prime} f_{k^\prime}^2 n_B(\omega_{k^\prime})}
%\sum_{N_\mathrm{out},N_\mathrm{in}}
%\sum_{\substack{k_1,\cdots,k_{N_\mathrm{out}},\\ k_1^\prime,\cdots,k^\prime_{N_\mathrm{in}}}}
%f_{k_1}^2 \cdots f_{k_{N_\mathrm{out}}}^2 f_{k_1^\prime}^2 \cdots f_{k^\prime_{N_\mathrm{in}}}^2
%(1+n_B (\omega_{k_1})) \cdots (1+n_B (\omega_{k_{N_\mathrm{out}}}))
%\times \nonumber
%\\ %\nonumber
%&	
%%\negthickspace\negthickspace\negthickspace\negthickspace
%%(1+n_B (\omega_{k_1})) \cdots (1+n_B (\omega_{k_{N_\mathrm{out}}}))
%\qquad\qquad
%n_B (\omega_{k^\prime_{1}}) \cdots n_B (\omega_{k^\prime_{N_\mathrm{in}}}) 2 \Re \left\{\int_{0}^{\infty} \mathrm{d} t \left(e^{i(\omega_k + \omega_{k^\prime_1} + \cdots + \omega_{k^\prime_{N_\mathrm{in}}} -\omega_{k_1} - \cdots -\omega_{k_{N_\mathrm{out}}})t} - \{\omega_k \to -\omega_k\}\right)\right\}.
%\end{align}
%Next, each form factor is assigned its corresponding exponential factor. We now push the sums over the momenta into the product:
\begin{align} \label{eqnS:ginel2_inst_2}
\Gamma^\mathrm{in}_k
%= &
%2 \lambda_0^2 f_k^2 e^{-\sum_{k^\prime} \left(\tilde{f}_{k^\prime}^2 + 2 f_{k^\prime}^2 n_B(\omega_{k^\prime})\right)}
%\Re \left\{\int_{0}^{\infty} \mathrm{d} t \left(e^{i\omega_k t} - e^{-i\omega_k t}\right)
%\sum_{N_\mathrm{out},N_\mathrm{in}} \frac{1}{N_\mathrm{out}!N_\mathrm{in}!} \times \right. \nonumber \\
%&
%\qquad\qquad \qquad\qquad \qquad \left(\sum_{k_1}f_{k_1}^2 (1+n_B (\omega_{k_1})) e^{-i\omega_{k_1} t}\right)
%\cdots
%\left(\sum_{k_{N_\mathrm{out}}}f_{k_{N_\mathrm{out}}}^2 (1+n_B (\omega_{k_{N_\mathrm{out}}})) e^{-i\omega_{k_{N_\mathrm{out}}} t}\right) \times \nonumber \\
%&
%\qquad\qquad \qquad\qquad \qquad  \left.
%\left(\sum_{k^\prime_1}f_{k^\prime_1}^2 n_B (\omega_{k^\prime_1}) e^{i\omega_{k^\prime_1} t}\right)
%\cdots
%\left(\sum_{k^\prime_{N_\mathrm{in}}}f_{k^\prime_{N_\mathrm{in}}}^2 n_B (\omega_{k^\prime_{N_\mathrm{in}}}) e^{i\omega_{k^\prime_{N_\mathrm{in}}} t}\right)\right\}
%\nonumber \\
= & 2 \lambda_0^2 f_k^2 e^{-\sum_{k^\prime} \left(\tilde{f}_{k^\prime}^2 + 2 f_{k^\prime}^2 n_B(\omega_{k^\prime})\right)}
\Re \left\{\int_{0}^{\infty} \mathrm{d} t \left(e^{i\omega_k t} - e^{-i\omega_k t}\right)
\sum_{N_\mathrm{out},N_\mathrm{in}} \frac{\sigma^{N_\mathrm{out}}_\mathrm{out}(t) \sigma^{N_\mathrm{in}}_\mathrm{in}(t)}{N_\mathrm{out}!N_\mathrm{in}!} \right\},
\end{align}
%We note that Eq.~\eqref{eqnS:ginel2_inst_1} may be retrieved from Eq.~\eqref{eqnS:ginel2_inst_2} by expanding the brackets and dropping terms with high orders of $f_k^2$.
%Let us denote the following:
where we introduced the notation
\begin{equation} \label{eqnS:sigma}
\sigma_\mathrm{in}(t) = \sum_{k^\prime}f_{k^\prime}^2 n_B (\omega_{k^\prime}) e^{i\omega_{k^\prime} t}, \qquad
\sigma_\mathrm{out}(t) = \sum_{k^\prime}f_{k^\prime}^2 \left(1 + n_B (\omega_{k^\prime})\right) e^{-i\omega_{k^\prime} t}.
\end{equation}
We recall that the sum involves only terms with odd $N_\mathrm{out}+N_\mathrm{in} \ge 3$. We decompose it to two sums --- one with odd $N_\mathrm{out}$ and even $N_\mathrm{in}$, and the other with even $N_\mathrm{out}$ and odd $N_\mathrm{in}$:
%\begin{align} \label{eqnS:sum_decomp}
%\sum_{N_\mathrm{out},N_\mathrm{in}} \frac{\sigma^{N_\mathrm{out}}_\mathrm{out}(t) \sigma^{N_\mathrm{in}}_\mathrm{in}(t)}{N_\mathrm{out}!N_\mathrm{in}!} &= \sum_{N_\mathrm{out}=1,3,\cdots}\frac{\sigma^{N_\mathrm{out}}_\mathrm{out}(t)}{N_\mathrm{out}!} \sum_{N_\mathrm{in}=2,4,\cdots}\frac{\sigma^{N_\mathrm{in}}_\mathrm{in}(t)}{N_\mathrm{in}!}
%+ \sum_{N_\mathrm{out}=2,4,\cdots}\frac{\sigma^{N_\mathrm{out}}_\mathrm{out}(t)}{N_\mathrm{out}!} \sum_{N_\mathrm{in}=1,3,\cdots}\frac{\sigma^{N_\mathrm{in}}_\mathrm{in}(t)}{N_\mathrm{in}!} \nonumber \\
%&= \sinh\left(\sigma_\mathrm{out}(t)\right) \left(\cosh\left(\sigma_\mathrm{in}(t)\right) - 1\right) + \sinh\left(\sigma_\mathrm{in}(t)\right) \left(\cosh\left(\sigma_\mathrm{out}(t)\right) - 1\right).
%\end{align}
\begin{align} \label{eqnS:sum_decomp}
\sum_{N_\mathrm{out},N_\mathrm{in}} \frac{\sigma^{N_\mathrm{out}}_\mathrm{out}(t) \sigma^{N_\mathrm{in}}_\mathrm{in}(t)}{N_\mathrm{out}!N_\mathrm{in}!} &= \hspace{0.33cm} \sum_{N_\mathrm{out}=1,3,\cdots}\frac{\sigma^{N_\mathrm{out}}_\mathrm{out}(t)}{N_\mathrm{out}!} \sum_{N_\mathrm{in}=2,4,\cdots}\frac{\sigma^{N_\mathrm{in}}_\mathrm{in}(t)}{N_\mathrm{in}!}
+
\sum_{N_\mathrm{out}=3,5,\cdots}\frac{\sigma^{N_\mathrm{out}}_\mathrm{out}(t)}{N_\mathrm{out}!}
\nonumber \\
&\hspace{0.4cm} + \sum_{N_\mathrm{out}=2,4,\cdots}\frac{\sigma^{N_\mathrm{out}}_\mathrm{out}(t)}{N_\mathrm{out}!} \sum_{N_\mathrm{in}=1,3,\cdots}\frac{\sigma^{N_\mathrm{in}}_\mathrm{in}(t)}{N_\mathrm{in}!} 
+
\sum_{N_\mathrm{in}=3,5,\cdots}\frac{\sigma^{N_\mathrm{in}}_\mathrm{in}(t)}{N_\mathrm{in}!}
\nonumber \\
&= \sinh\left[\sigma_\mathrm{out}(t) + \sigma_\mathrm{in}(t)\right] - \left[\sigma_\mathrm{out}(t) + \sigma_\mathrm{in}(t)\right].
\end{align}
We now plug this back to \eqref{eqnS:ginel2_inst_2} and combine the exponential factor outside of the integral with \eqref{eqnS:sum_decomp}. For the hyperbolic sine term we have:
\begin{align} \label{eqnS:sinh}
e^{-\sum_{k^\prime} \left(\tilde{f}_{k^\prime}^2 + 2 f_{k^\prime}^2 n_B(\omega_{k^\prime})\right)} \sinh\left(\sigma_\mathrm{out}(t) + \sigma_\mathrm{in}(t)\right) &= \nonumber \\
&\hspace{-3cm} \frac{1}{2} \exp \left(
-\sum_{k^\prime} \left\{  f_{k^\prime}^2 \left[ (1+n_B (\omega_{k^\prime}) ) (1 - e^{-i\omega_{k^\prime}t}) + n_B (\omega_{k^\prime}) (1 - e^{i\omega_{k^\prime}t})\right] + \tilde{f}_{k^\prime}^2 - f_{k^\prime}^2 \right\} \right) \nonumber \\
&\hspace{-3.4cm}- \frac{1}{2} \exp \left(
-\sum_{k^\prime} \left\{  f_{k^\prime}^2 \left[ (1+n_B (\omega_{k^\prime}) ) (1 + e^{-i\omega_{k^\prime}t}) + n_B (\omega_{k^\prime}) (1 + e^{i\omega_{k^\prime}t})\right] + \tilde{f}_{k^\prime}^2 - f_{k^\prime}^2 \right\} \right).
\end{align}
%so that each term is multiplied by $e^{-\sum_{k^\prime}f_{k^\prime}^2 n_B(\omega_{k^\prime})}$. For instance, for the hyperbolic sine of $\sigma_\mathrm{in}(t)$, we have:
%\begin{equation} \label{eqnS:sinh}
%e^{-\sum_{k^\prime}f_{k^\prime}^2 n_B(\omega_{k^\prime})}\sinh\left(\sigma_\mathrm{in}(t)\right) = \frac{1}{2}\left[\exp\left(-\sum_{k^\prime}f_{k^\prime}^2 n_B(\omega_{k^\prime})\left(1 - e^{i\omega_{k^\prime}t}\right)\right) - \exp\left(-\sum_{k^\prime}f_{k^\prime}^2 n_B(\omega_{k^\prime})\left(1 + e^{i\omega_{k^\prime}t}\right)\right)\right].
%\end{equation}
At low frequencies, one may approximate $f_q^2 \sim 2\Delta/z\omega_q$ [cf.~Eq.~\eqref{eqnS:f_k}]. The $1/\omega_q$ behavior leads to a logarithmic divergence of the sum over the modes, $\sum_{vq < \omega_c} f_q^2 \sim (2/z)\log(\Delta/2\omega_c)$, since the first mode is located at $\omega_q = \Delta/2$ (assuming open circuit boundary conditions at the far end of the line~\cite{kuzmin19b}). Here $\omega_c$ is a cutoff frequency which satisfies $\omega_0 - \omega_c \gg \Gamma_0$ and $\omega_c \gg \Gamma_0$. This logarithmic divergence is avoided in the first term on the right hand side in Eq.~\eqref{eqnS:sinh} due to the $\left(1 - e^{i\omega_{k^\prime}t}\right)$ factor, hence its contribution is finite. However, the second term goes to zero as a power law in the system size, $\sim \Delta^{2/z} \sim N^{-2/z}$. Thus, in the thermodynamic limit we may approximate the hyperbolic sine in \eqref{eqnS:sum_decomp} as an exponential, $\sinh [\sigma_\mathrm{out}(t) + \sigma_\mathrm{in}(t)] \approx e^{\sigma_\mathrm{out}(t) + \sigma_\mathrm{in}(t)}/2$, and neglect the linear term.
%(with a factor $1/2$). The contribution of the linear term in \eqref{eqnS:sum_decomp} vanishes due to the same reason.
%We then identify sums over odd $N$ as the Taylor expansions of hyperbolic sines, and sums over even $N$ as the Taylor expansions of hyperbolic cosines without the constant term. In the thermodynamic limit, the arguments of the hyperbolic functions are large, and the functions may be approximated as exponentials \textcolor{blue}{MAYBE ADD AN EQUATION}. We identify the sums as the Taylor series of exponentials.
This leads to $\Gamma^\mathrm{in}_k = 2f_k^2 \Im \Pi_R (\omega_k)$, with
%\begin{widetext}
\begin{equation} \label{eqnS:pi_inst}
\Pi_R(\omega) = -\lambda_0^2 %\tanh\frac{\omega}{2T}
%\left[
\int_0^\infty \mathrm{d}t
\sin(\omega t)
\exp \left(
-\sum_{k^\prime} \left\{  f_{k^\prime}^2 \left[ (1+n_B (\omega_{k^\prime}) ) (1 - e^{-i\omega_{k^\prime}t}) + n_B (\omega_{k^\prime}) (1 - e^{i\omega_{k^\prime}t})\right] + \tilde{f}_{k^\prime}^2 - f_{k^\prime}^2 \right\} \right), %\right],
\end{equation}
%\begin{equation} \label{eqnS:ginel2_inst_3}
%\Gamma^\mathrm{in}_k = 
%- 2 \lambda_0^2 f_k^2
%\Im \left\{\int_{0}^{\infty} \mathrm{d} t \sin(\omega_k t)
%\exp \left(
%-\sum_{k^\prime} \left\{  f_{k^\prime}^2 \left[ (1+n_B (\omega_{k^\prime}) ) (1 - e^{-i\omega_{k^\prime}t}) + n_B (\omega_{k^\prime}) (1 - e^{i\omega_{k^\prime}t})\right] + \tilde{f}_{k^\prime}^2 - f_{k^\prime}^2 \right\} \right)\right\},
%\end{equation}
which is Eq.~\eqref{eqn:pi_inst} of the main text. The on-resonance inelastic rate decreases with increasing $\omega_0/\Gamma_0$ due to the exponential decay of $\lambda_0$; dividing the rate by $\lambda_0^2$ reveals that the other factors in the last equation increase with increasing $\omega_0/\Gamma_0$, reflecting the increased phase space for inelastic processes, as depicted in Fig.~\ref{figS:comp}(a).

A similar procedure gives $\Gamma^\mathrm{in}_{k^\prime | k}$, the net rate of creation of photons at $k^\prime$ due to processes involving photons at $k$. Both $k$ and $k^\prime$ could correspond to either incoming or outgoing photons, and therefore $\Gamma^\mathrm{in}_{k^\prime | k}$ includes four terms. Let us start with the processes in which the photon at $k$ is absorbed and the photon at $k^\prime$ is emitted. Starting from Eq.~\eqref{eqnS:ginel2_inst_1}, we  use the delta function with $+\omega_k$ and separate out a factor $f_{k^\prime}^2(1 + n_B(\omega_{k^\prime}))$ from the sum over the outgoing photons. We thus have
\begin{align} \label{eqnS:ginel_kkp_1}
\Gamma^\mathrm{in~(1)}_{k^\prime | k} = &
\lambda_0^2 f_k^2 f_{k^\prime}^2(1 + n_B(\omega_{k^\prime})) e^{-\sum_{k^\prime} \left(\tilde{f}_{k^\prime}^2 + 2 f_{k^\prime}^2 n_B(\omega_{k^\prime})\right)}
\sum_{N_\mathrm{out},N_\mathrm{in}}
\sum_{\substack{k_1<\cdots<k_{N_\mathrm{out}},\\ k_1^\prime<\cdots<k^\prime_{N_\mathrm{in}}}}
f_{k_1}^2 \cdots f_{k_{N_\mathrm{out}}}^2 f_{k_1^\prime}^2 \cdots f_{k^\prime_{N_\mathrm{in}}}^2 n_B (\omega_{k^\prime_{1}}) \cdots n_B (\omega_{k^\prime_{N_\mathrm{in}}}) 
\times \nonumber
\\ %\nonumber
&	
%\negthickspace\negthickspace\negthickspace\negthickspace
%(1+n_B (\omega_{k_1})) \cdots (1+n_B (\omega_{k_{N_\mathrm{out}}}))
\qquad\qquad
(1+n_B (\omega_{k_1})) \cdots (1+n_B (\omega_{k_{N_\mathrm{out}}})) 
2\pi \delta \left( \omega_k - \omega_{k^\prime} + \omega_{k^\prime_1} + \cdots + \omega_{k^\prime_{N_\mathrm{in}}} -\omega_{k_1} - \cdots -\omega_{k_{N_\mathrm{out}}} \right),
\end{align}
where this time the sum is restricted to even $N_\mathrm{out}+N_\mathrm{in} \ge 2$.
Repeating the same procedure as before, we find
\begin{align} \label{eqnS:ginel_kkp_2}
\Gamma^\mathrm{in~(1)}_{k^\prime | k} =& 
\lambda_0^2 f_k^2 f_{k^\prime}^2(1 + n_B(\omega_{k^\prime}))
\Re \left\{\int_{0}^{\infty} \mathrm{d} t e^{i\left(\omega_k - \omega_{k^\prime}\right)t} \times \right. \nonumber \\
& \quad \left. \exp \left(
-\sum_{k^\prime} \left\{  f_{k^\prime}^2 \left[ (1+n_B (\omega_{k^\prime}) ) (1 - e^{-i\omega_{k^\prime}t}) + n_B (\omega_{k^\prime}) (1 - e^{i\omega_{k^\prime}t})\right] + \tilde{f}_{k^\prime}^2 - f_{k^\prime}^2 \right\} \right)\right\}.
\end{align}
The integral in Eq.~\eqref{eqnS:ginel_kkp_2} is very similar to the integral appearing in the self energy $\Pi_R(\omega)$ [Eq.~\eqref{eqnS:pi_inst}], but only accounts for absorption of a photon with energy $\omega_k - \omega_{k^\prime}$, whereas $\Pi_R(\omega)$ accounts for both emission and absorption processes. However, the emission rate (which includes an exponential $e^{-i(\omega_k - \omega_{k^\prime})t}$) is equal to the absorption rate times a factor $e^{-(\omega_k - \omega_{k^\prime})/T}$, as one may infer from the corresponding Bose-Einstein factors in Eq.~\eqref{eqnS:ginel2_inst_1}. Therefore, we find $\Gamma^\mathrm{in~(1)}_{k^\prime | k} = 2 f_k^2 f_{k^\prime}^2 (1 + n_B(\omega_{k^\prime}))(1 + n_B(\omega_k - \omega_{k^\prime})) \Im \Pi_R(\omega_k - \omega_{k^\prime})$. The remaining three terms of $\Gamma^\mathrm{in}_{k^\prime | k}$ may be derived in a similar fashion. This leads to Eq.~\eqref{eqn:ginel_kkp} of the main text:
\begin{align} \label{eqnS:ginel_kkp}
\Gamma^\mathrm{in}_{k^\prime | k} = 2f_k^2 f_{k^\prime}^2 & \left\{
\Im \Pi_R \left( \omega_k - \omega_{k^\prime} \right)
\left[ (1+n_B (\omega_{k^\prime})) (1+n_B (\omega_k - \omega_{k^\prime}))
- n_B (\omega_{k^\prime}) n_B (\omega_k - \omega_{k^\prime}) \right]
%\left( \coth \frac{\omega_k - \omega_{k^\prime}}{2T}  + \coth \frac{\omega_{k^\prime}}{2T} \right)
\right. \nonumber \\ & \left.
+ \Im \Pi_R \left( \omega_k + \omega_{k^\prime} \right)
\left[ (1+n_B (\omega_{k^\prime})) n_B (\omega_k + \omega_{k^\prime})
- n_B (\omega_{k^\prime}) (1+n_B (\omega_k + \omega_{k^\prime})) \right]
%\left( \coth \frac{\omega_k + \omega_{k^\prime}}{2T}  - \coth \frac{\omega_{k^\prime}}{2T} \right)
\right\}.
\end{align}
The rates $\Gamma^\mathrm{in}_k$ and $\Gamma^\mathrm{in}_{k^\prime | k}$ satisfy an energy conservation sum rule, which may be verified by integration by parts over $t$ in Eq.~\eqref{eqnS:pi_inst} for the self energies:
\begin{equation} \label{eqnS:energy_cons}
\omega_k \Gamma^\mathrm{in}_k = \sum_{k^\prime} \omega_{k^\prime} \Gamma^\mathrm{in}_{k^\prime | k}.
\end{equation}

\begin{figure*}
	\centering
	\includegraphics[width=0.45\columnwidth,height=!]{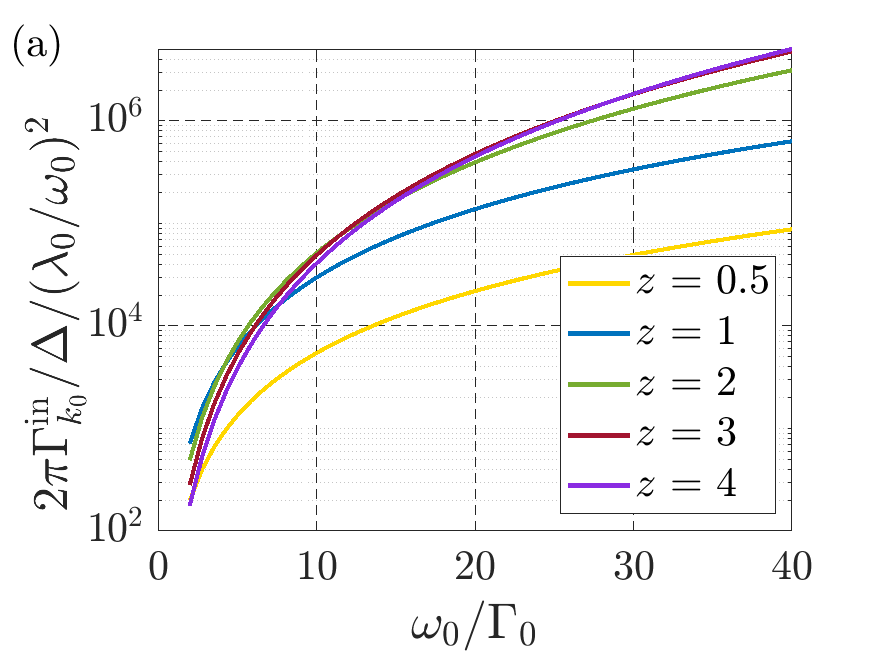}
	\includegraphics[width=0.45\columnwidth,height=!]{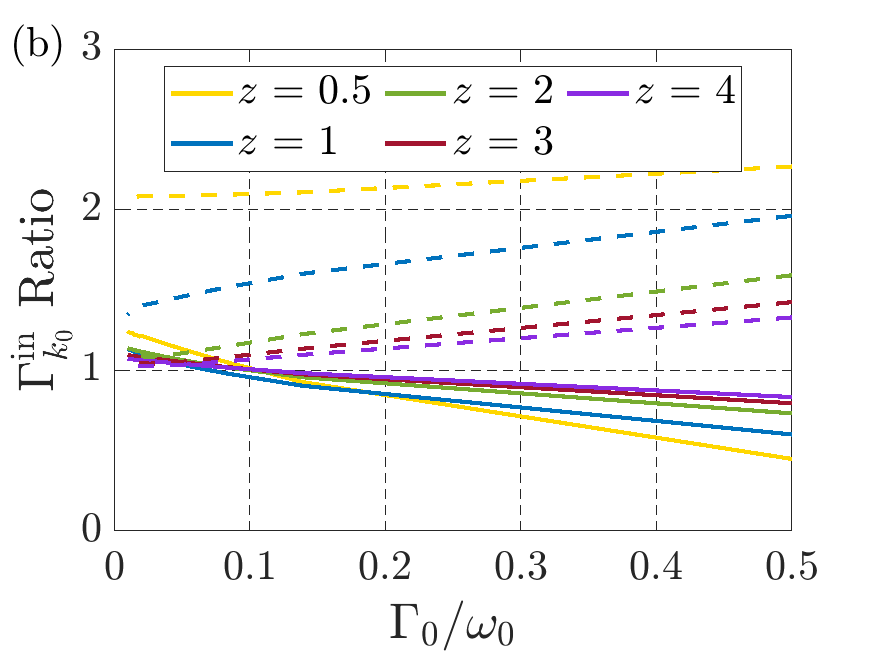}
	\caption{\label{figS:comp}
		(a) On-resonance (mode $k_0 = \omega_0/v$) total photon-instanton inelastic scattering probability $2\pi\Gamma^\mathrm{in}_{k_0}/\Delta$ [Eq.~\eqref{eqnS:pi_inst}] as function of $\omega_0/\Gamma_0$ for several values of $z$ at $T=0$,
		with the prefactor $\lambda_0^2$ excluded. This shows the increased phase space for scattering with increasing $\omega_0/\Gamma_0$.
		(b) Ratio between the $T=0$ on-resonance probabilities given by the instanton calculation, Eq.~\eqref{eqnS:pi_inst}, and either the dual cosine approximation, Eq.~\eqref{eqnS:sigma_ps} (continuous lines) or the first form of the limiting expression, Eq.~\eqref{eqnS:2f1} (dashed lines).
		%(c) uses fixed $z=2$, (e) uses fixed $\Gamma_0 = 0.05\omega_0$, and all other figures use a fixed value of $\Gamma_0 = 0.2\omega_0$	
	}
\end{figure*}

\subsection{Limiting cases} \label{subsec:limits}
To address limiting cases, it is useful to examine the behavior of $\Im \Pi_R(\omega)$ at low frequency, $\omega \ll \omega_0$.
Here we may use $f_q \sim \sqrt{2\Delta/z\omega_q}$ for $\omega_q \ll \omega_0$ [cf.~Eq.~\eqref{eqnS:f_k}] in Eq.~\eqref{eqnS:pi_inst} to obtain~\cite{gogolin} %the following approximation for the low-frequency behavior of $\Pi_R(\omega)$, based on :
%\begin{align} \label{eqnS:pi_inst_low}
%\Im \Pi_R(\omega) \approx \frac{\pi}{2\Gamma(2/z)} \frac{\lambda_0^2}{\omega}\left( \frac{\omega}{\omega_c(z)} \right)^{2/z},
%\quad \omega \ll \omega_0,
%\end{align}
\begin{align} \label{eqnS:pi_inst_low}
	\Im \Pi_R(\omega)
	\approx & \frac{1}{2\Gamma(2/z)} \frac{\lambda_0^2}{\omega_c(z)}\left( \frac{2\pi T}{\omega_c(z)} \right)^{2/z-1} \sinh\left(\frac{\omega}{2T}\right) \left| \Gamma \left( \frac{1}{z} + i \frac{\omega}{2\pi T}\right)\right|^2
	\nonumber \\
	\approx &
	\begin{cases}
		\frac{\pi}{2\Gamma(2/z)} \frac{\lambda_0^2}{\omega_c(z)}\left( \frac{\omega}{\omega_c(z)} \right)^{2/z-1}, & T \ll \omega \ll\omega_0, \\
		\frac{ \pi [\Gamma(1/z)]^2}{4\Gamma(2/z)} \frac{\lambda_0^2}{\omega_c(z)} \frac{\omega}{\omega_c(z)} \left( \frac{2 \pi T}{\omega_c(z)} \right)^{2/z-2}, & \omega \ll T \ll \omega_0,
	\end{cases}
\end{align}
where $\Gamma(x)$ is the gamma function~\cite{abramowitz}, and the effective cutoff $\omega_c(z) \approx 0.9 \omega_0$ is $z$ independent for $z \lesssim 1$.
The first case is Eq.~\eqref{eqn:pi_inst_low} of the main text.
We may thus immediately conclude that for $T=0$ and $\omega_k \ll \omega_0$, using $\Gamma^\mathrm{in}_k = 2f_k^2 \Im \Pi_R (\omega_k)$ with $f_k^2 \approx 2 \Delta/z\omega_k$ [cf.~Eq.~\eqref{eqnS:f_k}] gives
\begin{equation} \label{eqnS:ginel_T=0_low_low}
	\frac{2\pi\Gamma^\mathrm{in}_{k \ll k_0}}{\Delta} \approx \frac{4 \pi^2 (\omega_0/\omega_c)^{2/z}}{z \Gamma(2/z)} \left(\frac{\lambda_0}{\omega_0}\right)^2 \left(\frac{\omega_0}{\omega_k}\right)^{2 - 2/z}.
\end{equation}

\begin{figure*}
	\centering
	\includegraphics[width=0.34\columnwidth,height=!]{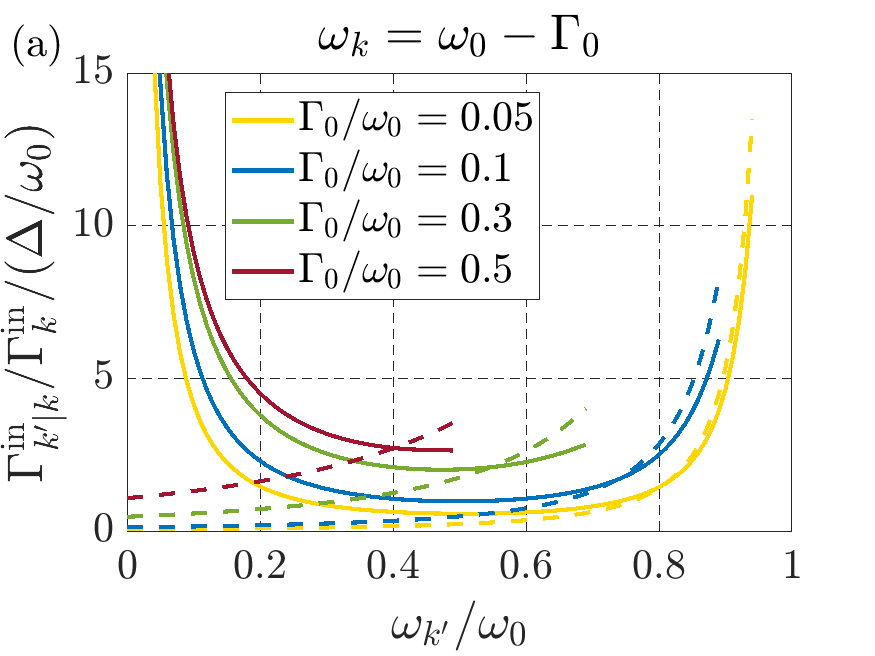}
	\hspace{-0.03\columnwidth}
	\includegraphics[width=0.34\columnwidth,height=!]{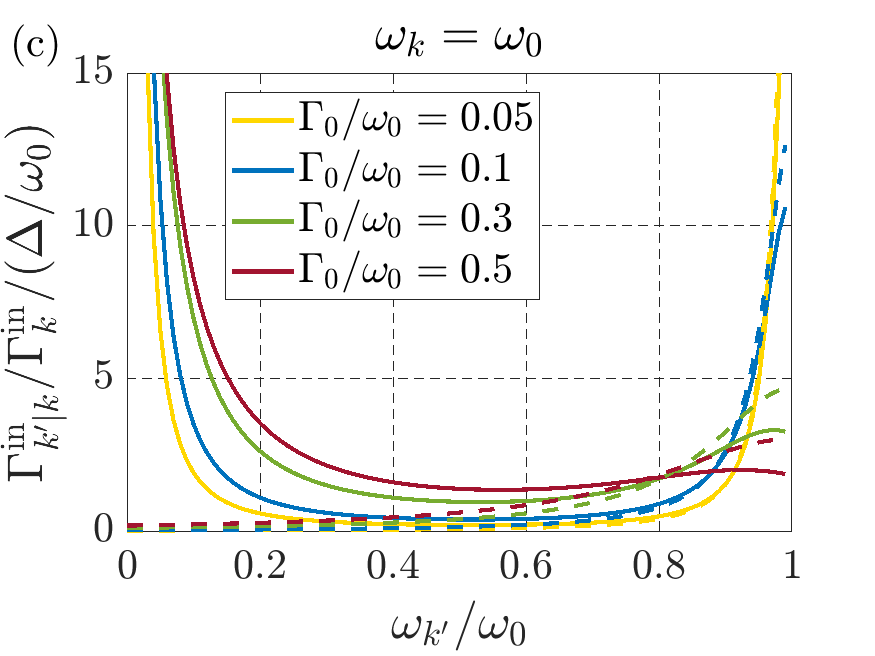}
	\hspace{-0.03\columnwidth}
	\includegraphics[width=0.34\columnwidth,height=!]{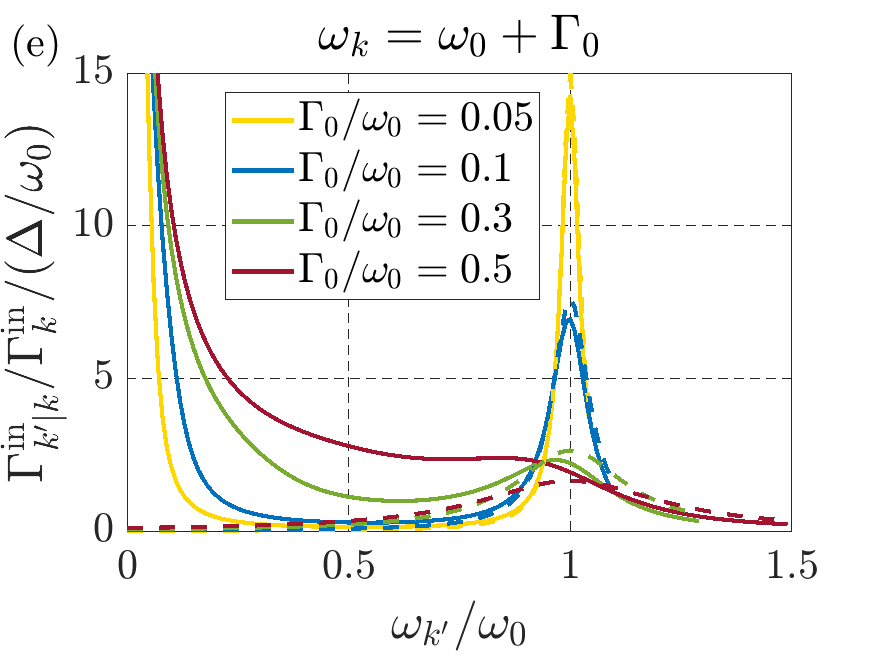}
	\newline
	\includegraphics[width=0.34\columnwidth,height=!]{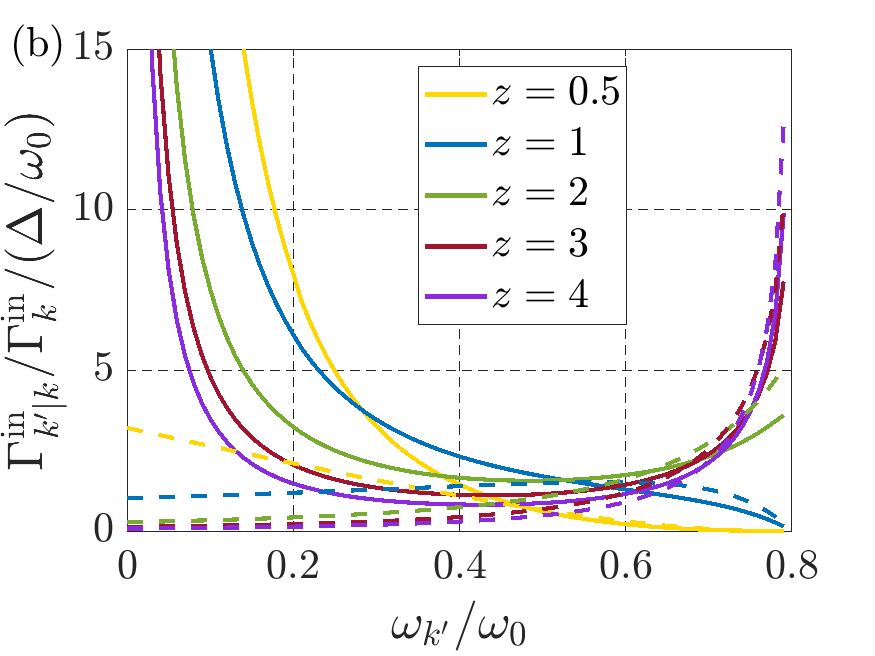} %{Gamma_inel_on_resonance_mathieu_wout_lambda_0.png}
	\hspace{-0.03\columnwidth}
	\includegraphics[width=0.34\columnwidth,height=!]{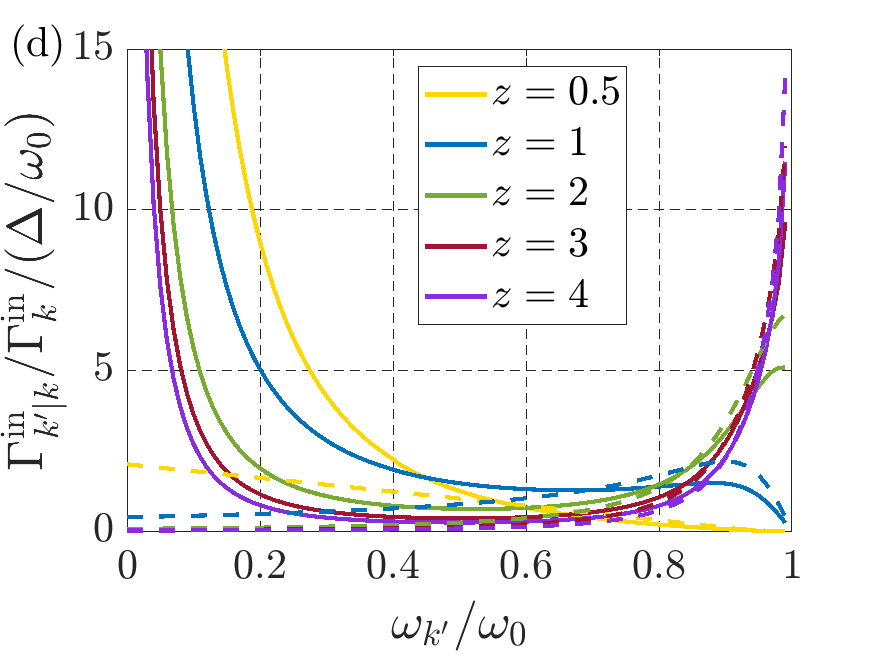}
	\hspace{-0.03\columnwidth}
	\includegraphics[width=0.34\columnwidth,height=!]{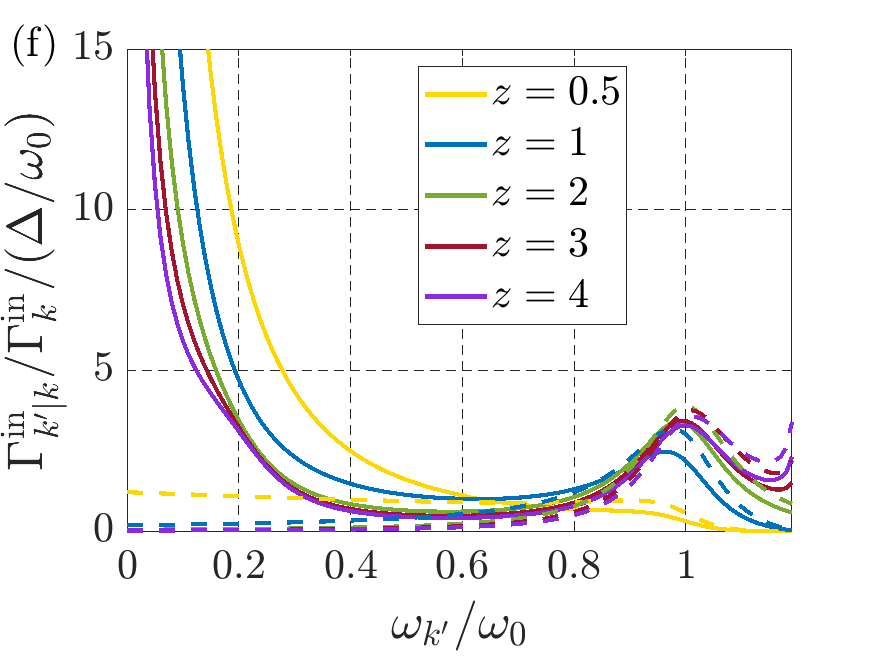}	
	\caption{\label{figS:inelastic_spectrum}
		%%(Color online)
		The distribution of inelastically generated photons $\Gamma^\mathrm{in}_{k^\prime|k}/\Gamma^\mathrm{in}_{k}$ at $T=0$ for (a,b) $\omega_k=\omega_0 - \Gamma_0$; (c,d) $\omega_k=\omega_0$; (e,f) $\omega_k=\omega_0 + \Gamma_0$. In panels (a), (c), and (e) $\Gamma_0/\omega_0$ varies at fixed $z=2$, while in panels (b), and (d), (f) $z$ varies at fixed $\Gamma_0/\omega_0=0.2$. Continuous lines represent the results of the full expressions, Eqs.~\eqref{eqnS:ginel_kkp} and~\eqref{eqnS:pi_inst}; dashed lines correspond to the approximation in Eq.~\eqref{eqnS:ginel_T=0_res_or_high}, and show that it applies for very small $\Gamma_0/\omega_0$ and/or large $z$.
	}
\end{figure*}

Let us now examine different limiting cases for the inelastic rate for $\omega_k$ comparable to $\omega_0$ and at $T=0$, where Eq.~\eqref{eqnS:ginel_kkp} reduces to $\Gamma^\mathrm{in}_{k^\prime | k} = 2f_k^2 f_{k^\prime}^2 \Im \Pi_R \left( \omega_k - \omega_{k^\prime} \right)$ [extensions to finite $T \ll \omega_0$ are possible based on the general form of Eq.~\eqref{eqnS:pi_inst_low}]. In the following, we assume $\Gamma_0/\omega_0 \to 0$ and $z > 1$.

%Using \eqref{eqnS:ginel_kkp} and the energy conservation sum rule, we have
%\begin{equation} \label{eqnS:ginel_T=0}
%\Gamma^\mathrm{in}_k = 2 f_k^2 \sum_{0 < k^\prime < k} \frac{\omega_{k^\prime}}{\omega_k} f_{k^\prime}^2 \Im \Pi_R \left( \omega_k - \omega_{k^\prime} \right).
%\end{equation}
%At $T=0$, the net rate of creation of photons at $k'$ due to processes involving photons at $k$ is given by
%\begin{equation} \label{eqnS:ginel_kkp}
%\Gamma^\mathrm{in}_{k^\prime | k} = 2f_k^2 f_{k^\prime}^2 \Im \Pi_R \left( \omega_k - \omega_{k^\prime} \right).
%\end{equation}
%The inelastic rate is related to $\Gamma^\mathrm{in}_{k^\prime | k}$ through the energy conservation sum rule:
%\begin{equation} \label{eqnS:energy_sum}
%\omega_k \Gamma^\mathrm{in}_k = \sum_{0 < k^\prime < k} \omega_{k^\prime} \Gamma^\mathrm{in}_{k^\prime | k}.
%\end{equation}
When either $|\omega_k - \omega_0| \lesssim \Gamma_0$ or $\omega_0 \gg \omega_k - \omega_0 \gg \Gamma_0$, %and $2\omega_0 - \omega_k \gg \Gamma_0$,
the main contribution to Eq.~\eqref{eqnS:energy_cons} would come from a near-resonance photon at $k^\prime$.
We may then approximate it as $\Gamma^\mathrm{in}_k = \sum_{k^\prime} \Gamma^\mathrm{in}_{k^\prime | k}$, and note that $f_q^2 \approx \Delta/\pi \left(\lambda_1/\lambda_0\right) (\Gamma_0/2)/[(\omega_0 - \omega_q)^2 + (\Gamma_0/2)^2]$ near resonance [here we use the WKB expressions for $\lambda_m$, cf.~Eq.~\eqref{eqnS:lambda_m}]. Furthermore, %both $\omega_k$ and $\omega_{k^\prime}$ are assumed to be close to resonance,
we may use Eq.~\eqref{eqnS:pi_inst_low} for $\Pi_R(\omega_k-\omega_{k^\prime})$. In addition, we introduce a cutoff $\omega_c^\prime$ [generally different than $\omega_c(z)$] which satisfies $\omega_c^\prime, \omega_0 - \omega_c^\prime \gg \Gamma_0$ as the lower summation limit, and evaluate the sum over $k^\prime$ as an integral:
\begin{equation} \label{eqnS:ginel_T=0_res_or_high}
\frac{ 2\pi\Gamma^\mathrm{in}_{k} }{\Delta} \approx \frac{2 \lambda_1^2 (\omega_0/\omega_c)^{2/z}}{\Gamma(2/z)} \frac{(\Gamma_0/2)}{(\omega_0 - \omega_k)^2 + (\Gamma_0/2)^2} \int_{\omega_c^\prime}^{\omega_k} \frac{\mathrm{d}\omega_{k^\prime}}{\omega_0} \left(\frac{\omega_k - \omega_{k^\prime}}{\omega_0}\right)^{2/z-1} \frac{(\Gamma_0/2)}{(\omega_0 - \omega_{k^\prime})^2 + (\Gamma_0/2)^2}.
\end{equation}
This is equivalent to Eq.~\eqref{eqn:ginel_kkp_knk0} in the main text.
As shown in Fig.~\ref{figS:inelastic_spectrum}, this is a good approximation for very small $\Gamma_0/\omega_0$ and/or large $z$.

For $|\omega_k - \omega_0| \lesssim \Gamma_0$ ($k$ close to $k_0=\omega_0/v$), we recognize that the leading $\omega_k$ dependence comes from the Lorentzian prefactor. Therefore, we set $\omega_k \approx \omega_0$ in the integrand, and replace the lower and upper integration limits by $0$ and $\omega_0$, respectively. Evaluation of the integral then leads to %Eq.~\eqref{eqn:2f1} of the main text:
\begin{align} \label{eqnS:ginel_knk0}
\negthickspace\negthickspace
%\frac{ \Gamma^\mathrm{in}_{k \sim k_0} }{\Gamma^\mathrm{in}_{k = k_0}}
\frac{ 2\pi\Gamma^\mathrm{in}_{k \sim k_0} }{\Delta}
\approx
\frac{ 2\pi\Gamma^\mathrm{in}_{k_0} }{\Delta}
\frac{(\Gamma_0/2)^2}{(\omega_0 - \omega_k)^2 + (\Gamma_0/2)^2},
\end{align}
with
\begin{align} \label{eqnS:2f1}
	%\negthickspace\negthickspace
	\frac{ 2\pi\Gamma^\mathrm{in}_{k_0} }{\Delta} \approx
	\frac{2 (\omega_0/\omega_c)^{2/z}}{\Gamma(1+2/z)} %\left(\frac{\omega_0}{\omega_c}\right)^{2/z}
	\frac{\lambda_1^2}{(\Gamma_0/2)^2} %\left(\frac{\lambda_1}{\Gamma}\right)^2
%	\frac{(\Gamma_0/2)^2}{(\omega_0 - \omega_k)^2 + (\Gamma_0/2)^2}
	{}_2F_1 \left(1, \frac{1}{z} ,\frac{z+1}{z}; -\frac{4\omega_0^2}{\Gamma_0^2} %-\left(\frac{\Gamma}{2\omega_0}\right)^2
	\right)
	\approx
	\frac{\pi (\omega_0/\omega_c)^{2/z}}{\Gamma(2/z)\sin(\pi/z)} %\left(\frac{\omega_0}{\omega_c}\right)^{2/z}
	\frac{\lambda_1^2}{(\Gamma_0/2)^2} %\left(\frac{\lambda_1}{\Gamma}\right)^2
%	\frac{(\Gamma_0/2)^2}{(\omega_0 - \omega_k)^2 + (\Gamma_0/2)^2}
\left(\frac{\Gamma_0/2}{\omega_0}\right)^{2/z},
\end{align}
where ${}_2F_1(a,b, c; x)$ is the hypergeometric function~\cite{abramowitz}. The last form relies again on $\Gamma_0/\omega_0 \ll 1$ and $z>1$, and corresponds to Eq.~\eqref{eqn:ginel_k0} in the main text.
The ratio between the on-resonance value of $\Gamma_k^\mathrm{in}$ evaluated using Eq.~\eqref{eqnS:pi_inst} and using the first form of Eq.~\eqref{eqnS:2f1} is displayed in Fig.~\ref{figS:comp}(b).

%For $\omega_k - \omega_0 \gg \Gamma_0$ and $2\omega_0 - \omega_k \gg \Gamma_0$,
For $\omega_0 \gg \omega_k - \omega_0 \gg \Gamma_0$
we start again from Eq.~\eqref{eqnS:ginel_T=0_res_or_high}. The leading contribution to the integral comes from $|\omega_{k^\prime} - \omega_0| \lesssim \Gamma_0$. Therefore, we set $\omega_{k^\prime} \approx \omega_0$ in the power-law factor and extend the integration limits to $\pm \infty$. This yields:
\begin{equation} \label{eqnS:ginel_kgk0} %\label{eqnS:ginel_T=0_high}
\frac{2 \pi \Gamma^\mathrm{in}_{k}}{\Delta} \approx \frac{2 \pi (\omega_0/\omega_c)^{2/z} }{\Gamma(2/z)} \frac{\lambda_1^2 \Gamma_0 / 2}{\omega_0^3} \left(\frac{\omega_0}{\omega_k - \omega_0}\right)^{3 - 2/z}.
\end{equation}

When $\omega_0 \gg \omega_0 - \omega_k \gg \Gamma_0$, %and $\omega_k \gg \Gamma_0$,
we return to Eq.~\eqref{eqnS:energy_cons}, drop the $(\Gamma_0 \omega_q)^2$ term in the denominator of $f_q^2$, and approximate the sines as linear functions for both $q=k,k^\prime$ [cf.~Eq.~\eqref{eqnS:f_k}]. We then use the low frequency approximation of $\Pi_R \left( \omega_k - \omega_{k^\prime} \right)$ [Eq.~\eqref{eqnS:pi_inst_low}], extend the lower integration limit to $-\infty$, and approximate $\omega_k/\omega_0 \approx 1$. We then find %Dropping a subleading term $(\omega_0/\omega_k)^2$, we find
\begin{equation} \label{eqnS:ginel_klk0} %\label{eqnS:ginel_T=0_low}
\frac{2\pi\Gamma^\mathrm{in}_{k}}{\Delta} \approx \frac{2 \pi (1 - 2/z) (\omega_0/\omega_c)^{2/z}} {\sin(2\pi/z) \Gamma(2/z)} \left(\frac{\lambda_1 \Gamma_0/2}{\omega_0^2}\right)^2 \left(\frac{\omega_0}{\omega_0 - \omega_k}\right)^{4 - 2/z}.
\end{equation}
Eqs.~\eqref{eqnS:ginel_knk0}, \eqref{eqnS:ginel_kgk0}, and \eqref{eqnS:ginel_kgk0} are summarized by Eq.~\eqref{eqn:ginel_resonance} in the main text.

\section{The dual cosine approach}
\label{sec:dual_cosine}
%\subsection{The inelastic rate}
The instanton approach presented in the previous section accounts for the full imaginary time dynamics of the phases along the array during a phase slip. A common phenomenological approach is to approximate a phase slip as an instantaneous $2\pi$ shift of all the phases ($\Phi_0=\pi$ shift of the fluxes) along the array. It has been successfully used in analyzing recent experiments on the system discussed~\cite{kuzmin20}. Here we will review this approach and show that, despite its crudeness, it gives rise to a reasonable approximation for the inelastic decay rate obtained by the full instanton calculation. %in the previous Section.

In a Hamiltonian formalism, sudden phase slips could be accounted for by supplementing the quadratic Hamiltonian corresponding to the Lagrangian~\eqref{eqnS:lag_harm} by a term~\cite{schon90,fazio01}
\begin{equation} \label{eqnS:hps}
H^{PS} = \lambda^{PS} \cos \left( \pi \sum_{n=0}^N Q_n \right)
= \lambda^{PS} \sum_{m=0}^\infty \frac{1}{(2m)!} \left[ \sum_q f^{PS}_q (a_q-a_q^\dagger)\right]^{2m},
%\lambda^{PS} \cosh \left[ \sum_k f^{PS}_k (a_k-a_k^\dagger)\right],
\end{equation}
%where the value of the coefficient $\lambda^{PS}$ will be discussed below, and
where $Q_0=C_0 \dot{\phi}_0$ and $Q_{n>0}=C_g \dot{\phi}_n$ denote the charge operators along the array, $a_q^\dagger$ ($a_q$) is the creation (annihilation) operator of a single photon in the array mode $q$, and where %the coefficients $f^{PS}_k$ are given by
\begin{equation} \label{eqnS:fk_ps}
f^{PS}_q = -i\sqrt{\frac{\omega_q}{2 C_q}}\left(C_0\sin\left(\delta_q\right) + C_g\sum_{n=1}^N\sin\left(qn + \delta_q\right)\right) = \sqrt{\frac{2\Delta}{z\omega_q}}
\frac{\omega_0^2}{ \left[ \left( \omega_q^2 - \omega_0^2 \right)^2 + (\Gamma_0 \omega_q)^2 \right]^{1/2}}.
\end{equation}
Comparing with Eq.~\eqref{eqnS:f_k} we can see that at low frequencies, $\omega_q \ll \omega_0$, $f^{PS}_q \approx f_q$, whereas near resonance they differ; at $k_0 = \omega_0/v$ we have $f^{PS}_{k_0} / f_{k_0} = \pi/4$.

The coefficient $\lambda^{PS}$ needs to be set by the value of a known observable. For the study of nearly-resonant photon scattering it is natural to choose it so that $H^{PS}$ reproduces the charge dispersion of the first excited level of an isolated transmon ($z \to \infty$ limit of our system), whose free Hamiltonian is $H_0 = \omega_0 a_0^\dagger a_0$. A phase slips operator corresponding to Eq.~\eqref{eqnS:hps} would be
\begin{equation} \label{eqnS:hps_iso}
	H^{PS}_{z\rightarrow \infty}(q_g) = \lambda^{PS} \cos \left( \pi(Q_0 - q_g) \right),
\end{equation}
where $Q_0 = (E_J/2E_C)^{1/4}(a_0 - a_0^\dagger)$ and $q_g$ is the gate charge offset. The first-order correction in perturbation theory to the first excited level is $E_1^{(1)}(q_g) = \langle H^{PS}_{z\rightarrow \infty}(q_g) \rangle {}_1$, and the half-width of the corresponding Bloch band is given by
\begin{equation} \label{eqnS:charge_disp_pert}
	\lambda_1 = \frac{E_1^{(1)}(q_g=1) - E_1^{(1)}(q_g=0)}{2} = -\lambda^{PS}\langle\cos(\pi Q_0)\rangle {}_1 = \pi^2 \lambda^{PS} \sqrt{ \frac{E_J}{2 E_C} } e^{-\pi^2 \sqrt{E_J / 8 E_C}}.
\end{equation}
%The expression to the left is equal to $\lambda_1$.
Within the WKB approximation in Eq.~\eqref{eqnS:lambda_m} we have $\lambda_1 = -\sqrt{2^7 E_J / E_C} \lambda_0$, hence $\lambda^{PS} = -(4/\pi)^2 e^{\pi^2 \sqrt{E_J / 8 E_C}} \lambda_0$.
%Thus, although $\lambda_0$ is exponentially small in $\sqrt{E_J/E_C}$, %with the exponential factor in Eq.~\eqref{eqnS:lambda_m}, one finds that
%$\lambda^{PS}$ is actually exponentially large. %grows exponentially with $E_J/E_C$.
%However, this exponential growth is countered by the sum over the resonant part of the form factor $f^{PS}_k$. This may actually be shown explicitly by pulling out a term $\sum_{k^\prime} \left(f_{k^\prime}^{PS}\right)^2$ from the exponential in Eq.~\eqref{eqnS:sigma_ps} and evaluating its contribution near resonance. We introduce a cutoff frequency $\omega_c$ on the order of $\omega_0$ which satisfies $\omega_0 \gg \omega_0 - \omega_c \gg \Gamma_0$.

Using the last form of Eq.~\eqref{eqnS:hps}, it is straightforward to calculate the self energy of a photon at mode $k$ to second order in $\lambda^{PS}$ as a sum of the contributions of the second-order self energies of each term in the sum over $m$ (including cross terms)~\cite{gogolin}. One factor of $f^{PS}_k (a_k-a_k^\dagger)$ is pulled out as an external leg in each term.
As usual, the calculation should be done in imaginary frequency space and analytically continued to real frequencies.
%Each external leg of the propagator fixes a single $q$ leg from Eq.~\eqref{eqnS:hps}, which leads to the following self-energy ($\left\langle\right\rangle_\omega$ denotes a Fourier transform):
We then find  $\Gamma^\mathrm{in}_k = 2f_k^2\Im \Pi^{PS}_R (\omega_k)$ with
\begin{align} \label{eqnS:sigma_ps}
\Pi^{PS}_R (\omega_k) &=
\sum_{m,m^\prime=0}^\infty \frac{i \left(\lambda^{PS}\right)^2}{(2m+1)!(2m^\prime+1)!} \int_0^\infty \mathrm{d}t e^{i\omega t} \left\langle \left[ \left\{ \sum_q f^{PS}_q \left( a_q e^{-i\omega_q t} -a_q^\dagger e^{+i\omega_q t} \right) \right\}^{2m+1}, \left\{ \sum_q f^{PS}_q \left( a_q -a_q^\dagger \right) \right\}^{2m^\prime+1} \right] \right\rangle \nonumber \\
& = i \left(\lambda^{PS}\right)^2 \int_0^\infty \mathrm{d}t e^{i\omega t} \left\langle \left[ \sin \left( \pi \sum_{n=0}^N Q_n(t) \right), \sin \left( \pi \sum_{n=0}^N Q_n(0) \right) \right] \right\rangle \nonumber \\
& = -\left(\lambda^{PS}\right)^2 \int_0^\infty \mathrm{d}t
\sin(\omega t)
\exp \left(
-\sum_{k^\prime} \left\{  \left(f^{PS}_{k^\prime}\right)^2 \left[ (1+n_B (\omega_{k^\prime}) ) (1 - e^{-i\omega_{k^\prime}t}) + n_B (\omega_{k^\prime}) (1 - e^{i\omega_{k^\prime}t})\right]\right\} \right).
\end{align}
%The inelastic rate is given by $\Gamma^\mathrm{in}_k = 2f_k^2\Im \Pi^{PS}(\omega_k)$.
We see that the result is very similar to the instanton inelastic rate, Eq.~\eqref{eqn:pi_inst} of the main text, with the replacements $\lambda_0 \to \lambda^{PS}$ and $f_k, \tilde{f}_k \to f_k^{PS}$. An analog of Eq.~\eqref{eqn:ginel_kkp} may  be derived as well, following similar steps to those in Sec.~\ref{ssec:ginel}, %~\ref{sec:instanton},
\begin{align} \label{eqnS:ginel_kkp_ps}
\Gamma^\mathrm{in}_{k^\prime | k} = 2\left(f^{PS}_k f^{PS}_{k^\prime}\right)^2 & \left\{
\Im \Pi^{PS}_R \left( \omega_k - \omega_{k^\prime} \right)
\left[ (1+n_B (\omega_{k^\prime})) (1+n_B (\omega_k - \omega_{k^\prime}))
- n_B (\omega_{k^\prime}) n_B (\omega_k - \omega_{k^\prime}) \right]
%\left( \coth \frac{\omega_k - \omega_{k^\prime}}{2T}  + \coth \frac{\omega_{k^\prime}}{2T} \right)
\right. \nonumber \\ & \left.
+ \Im \Pi^{PS}_R \left( \omega_k + \omega_{k^\prime} \right)
\left[ (1+n_B (\omega_{k^\prime})) n_B (\omega_k + \omega_{k^\prime})
- n_B (\omega_{k^\prime}) (1+n_B (\omega_k + \omega_{k^\prime})) \right]
%\left( \coth \frac{\omega_k + \omega_{k^\prime}}{2T}  - \coth \frac{\omega_{k^\prime}}{2T} \right)
\right\}.
\end{align}
%However, we still have to determine the prefactor $\lambda^{PS}$.

Interestingly, although $f_q^{PS}$ differs from $f_q, \tilde{f}_q$ near resonance, and $\lambda^{PS}$ differs from $\lambda_0$, the inelastic rates resulting from Eqs.~\eqref{eqnS:ginel_kkp} and~\eqref{eqnS:ginel_kkp_ps} are similar. Indeed, for $\omega_k, \omega_{k^\prime} \approx \omega_0$, when calculating $\Pi^{PS}_R \left( \omega_k - \omega_{k^\prime} \right)$ using Eq.~\eqref{eqnS:pi_inst}, only low-frequency modes contribute to the sum inside the exponent (see the discussion in Sec.~\ref{subsec:limits} above). Similar considerations apply to the last line of Eq.~\eqref{eqnS:sigma_ps} in this regime, except for a relative factor $e^{\sum_{k^{\prime\prime}} (f^{PS}_{k^{\prime\prime}})^2 - \tilde{f}_{k^{\prime\prime}}^2 }$, which is dominated by near resonance modes [at low frequencies $f^{PS}_q \approx f_q \approx \tilde{f}_q$, cf.~Eqs.~\eqref{eqnS:fk_ps}, \eqref{eqnS:f_k}, and~\eqref{eqnS:delta_S}].
Hence, collecting all the results of this section, the ratio of Eqs.~\eqref{eqnS:ginel_kkp_ps} and~\eqref{eqnS:ginel_kkp} is
\begin{align}
	\frac{ \left(f_k f_{k^\prime}\right)^2 \Im \Pi_R \left( \omega_k - \omega_{k^\prime} \right) }
	{\left(f^{PS}_k f^{PS}_{k^\prime}\right)^2 \Im \Pi^{PS}_R \left( \omega_k - \omega_{k^\prime} \right)}
	\approx &
	\left( \frac{f_{k_0}}{f^{PS}_{k_0}} \frac{f_{k_0}}{f^{PS}_{k_0}} \frac{\lambda_0}{\lambda^{PS}}\right)^2 e^{\sum_{k^{\prime\prime}} (f^{PS}_{k^{\prime\prime}})^2 - \tilde{f}_{k^{\prime\prime}}^2 }
	\\ \nonumber
	\approx &
	\left( \frac{4}{\pi} \frac{4}{\pi} \frac{\pi^2}{16} e^{- \pi^2 \sqrt{E_J / 8 E_C}} \right)^2 \exp \left[ \sum_{k^{\prime\prime}} \frac{\Delta\omega_0}{2z} \frac{1}{(\omega_0 - \omega_{k^{\prime\prime}})^2 + (\Gamma_0/2)^2 } \right]
	\approx 1.
\end{align}
Correspondingly, the ratio between the on-resonance instanton total inelastic rate [Eq.~\eqref{eqnS:pi_inst}] and the dual cosine total inelastic rate [Eq.~\eqref{eqnS:sigma_ps}] is close to unity, as shown in Fig.~\ref{figS:comp}(b).

\section{Derivation of the quartic nonlinearity inelastic rate}
\label{sec:quartic}
In this section we give some additional details regarding the derivation of the quartic nonlinearity inelastic rate, given by Eq.~\eqref{eqn:ginel_quartic} of the main text.
Here the quadratic Hamiltonian corresponding to the Lagrangian~\eqref{eqnS:lag_harm} is supplemented by the
quartic term in the Taylor expansion of the transmon Josephson cosine, $H^{(4)} = -2 E_J \phi_0^4/3$. The latter may be expressed in terms of the creation and annihilation operators of the eigenmodes ($a_k^\dagger$ and $a_k$, respectively, for mode $k$):
\begin{equation} \label{eqnS:hnl}
H^{(4)} = -\frac{2E_J}{3} \sum_{k_1, k_2, k_3, k_4} \prod_{i=1}^4 f^{(4)}_{k_i} \left(a_{k_i} + a_{k_i}^\dagger\right),
\end{equation}
where $f^{(4)}_k = \sin(\delta_k)/\sqrt{2 C_k \omega_k}$ are the form factors. Expansion of Eq.~\eqref{eqnS:hnl} yields terms with different numbers of creation and annihilation operators, which give rise to $1 \rightarrow 3$ (a single annihilation operator), $2 \rightarrow 2$ and $3 \rightarrow 1$ photon processes.
%\begin{align} \label{eqnS:hnl_exp}
%H^{(4)} = -\frac{2E_J}{3} \sum_{k_1, k_2, k_3, k_4} \left(\prod_{i=1}^4 f^{(4)}_k \right) \left[4 a_{k_1}^\dagger a_{k_2}^\dagger a_{k_3}^\dagger a_{k_4} + 6 a_{k_1}^\dagger a_{k_2}^\dagger a_{k_3} a_{k_4} + 4 a_{k_1}^\dagger a_{k_2} a_{k_3} a_{k_4}\right] = H^{(4)}_{1 \rightarrow 3} + H^{(4)}_{2 \rightarrow 2} + H^{(4)}_{3 \rightarrow 1}.
%\end{align}
The inelastic rate may then be calculated for each of these processes %three terms in Eq.~\eqref{eqnS:hnl_exp}
using the Fermi golden rule. Summing up the results gives
\begin{align}
\Gamma_k^{\mathrm{in}} =& \frac{4 z^2}{3 \pi} \frac{ \omega_0^4 \Delta^4}{\Gamma_0^2}
\frac{\sin^2 (\delta_{k}) }{\omega_{k}}
\sum_{k_1, k_2, k_3} %{k_1,k_2,k_3} &
\frac{\sin^2 (\delta_{k_1}) }{\omega_{k_1}} 
\frac{\sin^2 (\delta_{k_2}) }{\omega_{k_2}}
\frac{\sin^2 (\delta_{k_3}) }{\omega_{k_3}}
\delta(\omega_k + \omega_{k_1} - \omega_{k_2} - \omega_{k_3})
\times \\ \nonumber
& \hspace{-0.5cm} \left[\left(1 + n_B(\omega_{k_1})\right) \left(1 + n_B(\omega_{k_2})\right) \left(1 + n_B(\omega_{k_3})\right) + 
n_B(\omega_{k_1}) \left(1 + n_B(\omega_{k_2})\right) \left(1 + n_B(\omega_{k_3})\right) + 
n_B(\omega_{k_1}) n_B(\omega_{k_2}) \left(1 + n_B(\omega_{k_3})\right)\right]
%\prod_{\tilde{k}=k,k_1,k_2,k_3} \frac{\sin^2 (\delta_{\tilde{k}}) }{\omega_{\tilde{k}}}
,
\end{align}
where the three terms in the square brackets on the last line correspond to $1 \rightarrow 3$, $2 \rightarrow 2$ and $3 \rightarrow 1$ scattering, respectively. At $T=0$ only the $1 \rightarrow 3$ process contributes,
%However, at low temperatures, only the $1 \rightarrow 3$ term contributes; to see that, consider for example the $2 \rightarrow 2$ term. Using the appropriate Bose-Einstein factors, we find:
%\begin{align}
%\Gamma^{2 \rightarrow 2}_{k} = \frac{4 z^2}{3 \pi} \frac{ \omega_0^4 \Delta^4}{\Gamma_0^2}
%\frac{\sin^2 (\delta_{k}) }{\omega_{k}}
%\sum_{k_1, k_2, k_3} %{k_1,k_2,k_3}
%\frac{\sin^2 (\delta_{k_1}) }{\omega_{k_1}} 
%\frac{\sin^2 (\delta_{k_2}) }{\omega_{k_2}}
%\frac{\sin^2 (\delta_{k_3}) }{\omega_{k_3}}
%n_B(\omega_{k_1}) \left(1 + n_B(\omega_{k_2})\right) \left(1 + n_B(\omega_{k_3})\right)
%%\prod_{\tilde{k}=k,k_1,k_2,k_3} \frac{\sin^2 (\delta_{\tilde{k}}) }{\omega_{\tilde{k}}}
%\delta(\omega_k + \omega_{k_1} - \omega_{k_2} - \omega_{k_3}).
%\end{align}
%The incoming thermal photons of $2 \rightarrow 2$ and $3 \rightarrow 1$ processes introduce Bose-Einstein factors which suppress the inelastic rate significantly at $T \ll \omega_k$. Therefore, we consider only the $H^{(4)}_{1 \rightarrow 3}$ term, which does not envolve thermal photons. We focus on $T = 0$; using the Fermi golden rule, we arrive at
\begin{align}
\Gamma^\mathrm{in}_{k} = \frac{4 z^2}{3 \pi}
\frac{ \omega_0^4 \Delta^4}{\Gamma_0^2}
\frac{\sin^2 (\delta_{k}) }{\omega_{k}}
\sum_{k_1, k_2, k_3} %{k_1,k_2,k_3}
\frac{\sin^2 (\delta_{k_1}) }{\omega_{k_1}} 
\frac{\sin^2 (\delta_{k_2}) }{\omega_{k_2}}
\frac{\sin^2 (\delta_{k_3}) }{\omega_{k_3}}
%\prod_{\tilde{k}=k,k_1,k_2,k_3} \frac{\sin^2 (\delta_{\tilde{k}}) }{\omega_{\tilde{k}}}
\delta(\omega_k - \omega_{k_1} - \omega_{k_2} - \omega_{k_3}),
\end{align}
which is Eq.~\eqref{eqn:ginel_quartic} of the main text. Turning the sums into integrals, and accounting for the delta function, one finds
\begin{align} \label{eqnS:gamma_quartic_integrals}
\Gamma_k^{\mathrm{in}} &= \frac{4 \Delta z^2 \omega_0^4 \Gamma_0^6}{3 \pi} \frac{\omega_k}{\left(\omega_0^2 - \omega_k^2\right)^2 + (\Gamma_0 \omega_k)^2} \int_{0}^{\omega_k}\mathrm{d} \omega_1 \int_{0}^{\omega_k - \omega_1} \mathrm{d} \omega_2 \nonumber \\
&\frac{\omega_1}{\left(\omega_0^2 - \omega_1^2\right)^2 + \left(\Gamma_0 \omega_1\right)^2} \frac{\omega_2}{\left(\omega_0^2 - \omega_2^2\right)^2 + \left(\Gamma_0 \omega_2\right)^2}
\frac{\omega_k - \omega_1 - \omega_2}{\left(\omega_0^2 - (\omega_k - \omega_1 - \omega_2)^2\right)^2 + \left(\Gamma_0(\omega_k - \omega_1 - \omega_2)\right)^2}.
\end{align}
The inelastic rate may be evaluated explicitly for a nearly-resonant incoming photon, $|\omega_k-\omega_0| \lesssim \Gamma_0$  ($k$ close to $k_0=\omega_0/v$). In this case the outgoing photons are all far from resonance. Hence, we may approximate the form factors inside the integrals by dropping the terms proportional to $\Gamma_0^2$ in the denominators. The form factor of the incoming photon may be approximated by a Lorentzian with width $\Gamma_0/2$. The dominant part of the $\omega_k$ dependence is captured by this Lorentzian, hence we replace $\omega_k \approx \omega_0$ inside the integrals. The inelastic rate then becomes
\begin{align} \label{eqnS:ginel_nl_near_resonance_1}
\frac{2\pi\Gamma_{k \sim k_0}^{\mathrm{in}}}{\Delta} \approx \frac{128 c_0 z^2}{3} \left(\frac{\Gamma_0/2}{\omega_0}\right)^4 \frac{(\Gamma_0/2)^2}{(\omega_0 - \omega_k)^2 + (\Gamma_0/2)^2},
\end{align}
with
\begin{align} \label{eqnS:ginel_nl_near_resonance_int}
c_0 &= \int_{0}^{1}\mathrm{d} x_1 \int_{0}^{1 - x_1} \mathrm{d} x_2
\frac{x_1}{\left(1 - x_1^2\right)^2} \frac{x_2}{\left(1 - x_2^2\right)^2}
\frac{1 - x_1 - x_2}{\left(1 - (1 - x_1 - x_2)^2\right)^2} \approx 0.026.
%\nonumber \\
%&= \frac{1}{2^{14}}\left(816 - 15\pi^2 - 688\log 2 - 90\log^2 2 + 72 \log 486 - 180\left(\mathrm{Li}_2 \left(-\frac{1}{2}\right) - \mathrm{Li}_2 \left(\frac{1}{3}\right) + \mathrm{Li}_2 \left(\frac{2}{3}\right)\right)\right) \nonumber \\
%&\approx 0.026,
\end{align}
%where $\mathrm{Li}_s(z)$ is the polylogarithm function.
%\newsavebox\mytempbib
%\savebox\mytempbib{\parbox{\textwidth}{\input{paper.bbl}}}
%\printbibliography\relax
%\addtocounter{framenumber}{-1}
%\begin{frame}<0>
%\input{paper.bbl}
%\end{frame}

\end{widetext}

%% file: large_EJ_paper_arxiv.bbl
\begin{thebibliography}{48}
\bibitem [{fn:({\natexlab{a}})}]{fn:DCB}%
  \BibitemOpen
  \bibinfo {note} {By the dynamical Coulomb blockade~\cite{ingold92}, inelastic
  Cooper pair tunneling (dual to phase slips) could also induce photon
  emission~\cite[][and references therein]{peugeot20}. However, that process is
  driven by a dc voltage bias, akin (by the ac Josephson effect) to strong
  pumping by a microwave source in our setup, in marked distinction from the
  single photon excitation regime we consider}\BibitemShut {NoStop}%
\bibitem [{fn:({\natexlab{b}})}]{fn:fluctuations}%
  \BibitemOpen
  \bibinfo {note} {Corrections to the fluctuations contribution are subleading
  in $\Gamma_0/\omega_0$}\BibitemShut {NoStop}%
\bibitem [{fn:({\natexlab{c}})}]{fn:thermodynamic}%
  \BibitemOpen
  \bibinfo {note} {The sum is restricted to odd $N_\mathrm{out} + N_\mathrm{in}
  \ge 3$. However, in the thermodynamic limit one may discard this restriction
  and introduce the prefactor of half~\cite[][Sec.~SI.D]{SM}}\BibitemShut
  {NoStop}%
\end{thebibliography}
